\documentclass[natbib209]{emulateapj}

\usepackage{amsmath}
\usepackage{epsfig}

\newcommand{\pd}{\partial}

\newcommand{\bdot}{\mbox{\boldmath $\cdot$}}
\newcommand{\del}{\mbox{\boldmath $\nabla$}}
\newcommand{\curl}{\mbox{\boldmath $\nabla \times$}}
\newcommand{\dv}{\mbox{\boldmath $\nabla \bdot$}}

\newcommand{\vv}{{\bf v}}

\newcommand{\vort}{\mbox{\boldmath $\omega$}}
\newcommand{\rh}{\overline{\rho}}

\newcommand{\ssurf}{\mbox{\boldmath ${\cal S}$}}

\newcommand{\uvr}{\mbox{\boldmath $\hat{r}$}}
\newcommand{\uvt}{\mbox{\boldmath $\hat{\theta}$}}

\newcommand{\uvn}{\mbox{\boldmath $\hat{n}$}}

\shorttitle{Convective Velocities in the Solar Interior}

\shortauthors{Miesch et al.}

\begin{document}

\title{On the Amplitude of Convective Velocities in the Deep Solar Interior}

\author{Mark S. Miesch$^1$, Nicholas A. Featherstone$^1$, Matthias Rempel$^1$ and
Regner Trampedach$^2$}
\affil{$^1$High Altitude Observatory, National Center for Atmospheric Research, Boulder, CO, 80307-3000, USA: miesch@ucar.edu}
\affil{$^2$JILA, University of Colorado and National Institute of Standards and Technology, 440 UCB, Boulder, CO 80309}

\begin{abstract}
We obtain lower limits on the amplitude of convective velocities in
the deep solar convection zone based only on the observed properties
of the differential rotation and meridional circulation together with
simple and robust dynamical balances obtained from the fundamental MHD
equations.  The linchpin of the approach is the concept of gyroscopic
pumping whereby the meridional circulation across isosurfaces of
specific angular momentum is linked to the angular momentum transport
by the convective Reynolds stress.  We find that the amplitude of the 
convective velocity must be at least 30 m s$^{-1}$ in the
upper CZ ($r \sim 0.95 R$) and at least 8 m s$^{-1}$ in the lower CZ
($r \sim 0.75 R$) in order to be consistent with the observed mean
flows.  Using the base of the near-surface shear layer as a probe of
the rotational influence, we are further able to show that the
characteristic length scale of deep convective motions must be no
smaller than 5.5--30 Mm.  These results are compatible with convection
models but suggest that the efficiency of the turbulent transport
assumed in advection-dominated flux-transport dynamo models is
generally not consistent with the mean flows they employ.
\end{abstract}

\section{Introduction}\label{sec:intro}

Our modern understanding of solar internal dynamics rests heavily on 
global and local helioseismology, with perspective provided by theoretical
and numerical models.  This perspective is particularly important in the
deepest regions of the convection zone where mean flows are established
and where helioseismic probing is most challenging.

The most reliable result from helioseismology with regard to the dynamics
of the deep convection zone (CZ) continues to be the solar internal rotation 
profile obtained from global inversions \citep{thomp03}. Such inversions 
indicate that the monotonic surface differential rotation ($\sim 30$\% decreases in angular
velocity $\Omega$ from equator to pole) persists throughout the convection
zone with little radial variation.  Further dynamical clues come from 
estimates of the meridional flow near the surface inferred from local 
helioseismic inversions and photospheric observations 
\citep{gonza06,basu10,ulric10,hatha10,hatha11b}.  Although various 
measurement techniques can yield disparate results for the radial 
and temporal dependence of the meridional flow, all techniques generally
agree that there is a persistent poleward flow near the surface 
($r \gtrsim 0.95 R$, where $R$ is the solar radius) from the equator 
up to latitudes of at least 60-70$^\circ$, with an amplitude of 
10-20 m s$^{-1}$.

Less is known about the structure or amplitude of convective flows in the 
deep convection zone.  Photospheric observations are dominated by solar
granulation, with a velocity scale of $\sim$ 2 km s$^{-1}$ and a size
scale of $\sim $ 1Mm, but deep convective motions are thought
to be much slower and larger \citep[e.g.][]{miesc09}.  Closer scrutiny of the 
photospheric velocity power spectrum obtained from Doppler measurements
shows a prominent peak near a spherical harmonic degree $\ell \sim 120$, 
corresponding to a size scale of 30-35 Mm and a spectral velocity amplitude 
of 7-8 m s$^{-1}$ \citep{hatha00}.  However, this observed
spectral amplitude is significantly lower than the true amplitude
of the convection due mainly to projection effects; the Doppler 
measurements trace mainly horizontal flows near the limb.
Taking projection and other sampling effects into account 
(e.g. spatial and temporal filtering), Hathaway et al. were
able to match the observed spectrum near $\ell \sim 120$
with a supergranular flow component having a typical velocity
amplitude of about 300-400 m s$^{-1}$, in agreement with
the correlation tracking of surface features \citep{deros04,roudi12}.
The power declines steadily toward lower $\ell$, where one would 
expect to see signatures of deep convection (giant cells).  Thus, 
either giant cells have a lower amplitude than supergranulation, 
or they do not imprint through to the photosphere (or both).

\cite{hanas10,hanas12} have recently searched for signatures of
subsurface convection in local helioseismic inversions at depths
of 0.92$R$, 0.95$R$, and 0.96$R$.  Their analysis indicates
that the spectral velocity amplitude at these depths is less than
10 m s$^{-1}$ and possibly less than 1 m s$^{-1}$ for convective
motions with $\ell < 60$ and correlation times longer than 96 hours.  
This is one to two orders of magnitude lower than suggested by 
convection simulations and mixing length theory 
(see \S\ref{sec:models}).

In this paper we argue that these results, if confirmed, would have
important implications for the maintnance of mean flows in the Sun.
In particular, we use fundamental physical arguments together with
solar observations to derive lower bounds on the amplitude of
convective motions in the deep solar interior.  These estimates are
based on robust dynamical balances deduced from the equations of
magnetohydrodynamics (MHD).  They are not based on mixing
length theory and they do not depend on any results from numerical
simulations, although they are consistent with both
(\S\ref{sec:models}).  We find that the amplitude of convective
motions in the upper convection zone ($r \sim 0.95$) must be at least
30 m s$^{-1}$ in order to be capable of sustaining the mean flows
inferred from helioseismology.  Similar estimates for the lower CZ ($r
\sim 0.75 R$) imply that convective motions there must be at least 8
m s$^{-1}$.  Reconciling these lower limits with the upper limits 
found by \citet{hanas10,hanas12} is a challenge for global convection 
models but may be possible if the motions responsible for maintaining
mean flows span multiple scales, with significant power above
$\ell \sim 60$.

In \S\ref{sec:meanflows} we discuss how differential rotation and meridional
circulation are maintained in the Sun by means of the convective Reynolds stress,
baroclinicity, and their own inertia.  This provides a link between convection
and mean flows that we exploit in \S\ref{sec:amp} to obtain lower limits
on the amplitude of convective motions in the deep convection zone.  
In \S\ref{sec:discussion} we address the length scale of convective motions
and the implications of our velocity and length scale estimates for 
helioseimic probing, flux-transport dynamo models, and convection
models.  Section \ref{sec:summary} is a summary of our principle
results and conclusions.

\section{Convective Origins of Mean Flows}\label{sec:meanflows}

\subsection{Gyroscopic Pumping}\label{sec:gp}

The conservation of angular momentum for a statistically steady 
flow in a rotating spherical shell may be expressed
as follows \citep[][hereafter MH11]{miesc11}:
\begin{equation}\label{eq:gp}
\left<\rho \vv_m\right> \bdot \del {\cal L} = {\cal F} ~~~. 
\end{equation}
where $\rho$ is the mass density, ${\cal L} = \lambda^2 \Omega$ is
the specific angular momentum, $\Omega$ is the rotation rate,
$\lambda = r\sin\theta$ is the cylindrical radius, and
$\left<\vv_m\right> = \left<v_r\right> \uvr + \left<v_\theta\right> \uvt$
is the mean meridional flow.  This equation is derived by averaging the zonal 
momentum equation over longitude and time, with averages denoted by angular 
brackets $\left< \right>$.  Spherical polar coordinates ($r$, $\theta$, $\phi$)
are used throughout.

The ${\cal F}$ term in eq.\ (\ref{eq:gp}) is a net axial torque 
that includes contributions from the Reynolds stress, 
the Lorentz force, and the viscous diffusion.  Explicit
expressions are given in MH11.  The molecular viscosity 
is small in stars so viscous diffusion can be safely neglected.  
Neglecting the Lorentz force is less justified.  However, the magnetic 
pressure gradient averages out when taking the mean so the only contribution
of the Lorentz force to ${\cal F}$ is from magnetic tension.  This
generally tends to suppress rotational shear \citep[e.g.][]{brun04}
so it is unlikely to be the dominant factor in establishing the
solar differential rotation.  In this paper we wish to establish 
a lower limit on how strong the Reynolds stress must be in order 
to balance the advection of angular momentum by the meridional flow; 
that is, the term on the left-hand-side (LHS) of (\ref{eq:gp}).  
If it must also balance magnetic tension, then the Reynolds stress
must be even stronger.  Therefore, the inclusion of the Lorentz force is 
not likely to change the lower bounds we establish in \S\ref{sec:amp}.

Thus, if we assume the Reynolds stress is the dominant contribution
to ${\cal F}$, Eq.\ (\ref{eq:gp}) becomes
\begin{equation}\label{eq:bob}
\left<\rho \vv_m\right> \bdot \del {\cal L} = - \dv \left( \left<\rho \lambda \vv_m^\prime v_\phi^\prime\right>\right)  ~~~.
\end{equation}
where primes indicate fluctuations about the mean.  Since $\del {\cal L}$
is cylindrically outward in the Sun, the implication here is that the 
angular momentum transport by the Reynolds stress (RHS) establishes
the meridional flow by inducing a flow toward the rotation axis
when the RHS is negative (divergence) and away from the rotation axis when
the RHS is positive (convergence).  This mechanism is known as gyroscopic pumping 
\citep[MH11;][]{mcint98} and is supported by convection simulations and
mean-field models \citep{rempe05,brun11,feath12}.

The fundamental physics behind gyroscopic pumping is discussed by
MH11, \cite{hayne91} and \cite{mcint98}.  In short, an axial
variation in the net torque $\pd {\cal F}/\pd z$ establishes an axial
shear $\pd \Omega / \pd z$ which in turn induces a meridional flow
through the Coriolis force (\S\ref{sec:twb}).  In order to determine
the nature of the differential rotation that will ultimately be
established, one must consider the meridional force balance as we do
in the next section.  However, the equilibrium structure of the
meridional flow is surprisingly less sensitive to the meridional force
balance and is instead regulated mainly by eq.\ (\ref{eq:gp}).  This 
is because the primary contribution to $\del {\cal L}$ is from the mean
(globally averaged) rotation rate of the Sun, $\Omega_0$ (see MH11 and
\S\ref{sec:estimates} below), so changes in the differential rotation,
$\del \Omega$, do not change the balance in eq.\ (\ref{eq:gp})
appreciably.

\subsection{Meridional Force Balance and the Role of Baroclinicity}\label{sec:twb}

In \S\ref{sec:gp} we argued that the mean meridional circulation in the solar
envelope is established and maintained through gyroscopic pumping, as
expressed by eq.\ (\ref{eq:gp}).
Since the process of gyroscopic pumping is mediated by the Coriolis force,
this is equivalent to saying that the meridional circulation is maintained
by the inertia of the differential rotation (MH11).  However, it is well 
known that thermal gradients can also establish meridional circulation
by means of baroclinicity and it is sometimes argued that the meridional
flow in the solar convection zone may be baroclinic in nature.  In this
section we argue that this is not the case.  In particular, we consider 
the meridional force balance in the solar convection zone within the
context of solar observations and we argue that gyroscopic pumping, 
as expressed in eq.\ (\ref{eq:gp}), rather than baroclinicity, is the 
primary mechanism by which the solar meridional circulation is maintained.

The ambiguity arises because the thermal energy equation can be expressed
in a manner analogous to eq.\ (\ref{eq:gp}):
\begin{equation}\label{eq:therm}
\left<\rho \vv_m\right> \bdot \del \left<S\right> = {\cal Q} ~~~. 
\end{equation}
where ${\cal Q}$ involves the (negative) divergence of the convective 
entropy flux, the radiative diffusion, and the viscous and ohmic heating,
although the latter are negligible in stellar interiors.  

It is clear from equations (\ref{eq:gp}) and (\ref{eq:therm}) that both
mechanical and thermal forcing, ${\cal F}$ and ${\cal Q}$, can induce 
a meridional circulation.  Indeed, this is a classical problem in the
theory of planetary and stellar atmospheres 
\citep{elias51,read86,tasso78}.  In a solar context,
eq.\ (\ref{eq:therm}) plays an important role in the radiative 
spreading of the solar tachocline \citep{spieg92}.

We can be reasonably certain from solar observations that both 
${\cal F}$ and ${\cal Q}$ are nonzero.  If this were not the 
case, then a steady state would only be possible if either 
${\cal L}$ or $\left<S\right>$ were constant on streamlines 
of the meridional mass flux.  Helioseismic rotational inversions
together with observations of poleward mass flux in the solar
surface layers rule out the former (${\cal L}$ constant on 
streamlines).  The latter is ruled out by mass conservation
($\left<v_r\right>$ must be nonzero in the deep convection 
zone to sustain the poleward mass flux in the surface layers)
and the requirement that at least some portion of the convection
zone be superadiabatic $\pd \left<S\right>/\pd r < 0$.  

The relative contribution of mechanical and thermal forcing 
to establishing the meridional flow becomes more clear if
we consider the mean zonal vorticity equation 
\citep[e.g.][MH11]{miesc05,balbu09}.  This is obtained 
from the meridional components of the MHD momentum equation,
averaged over longitude and time.  Previous work based on
numerical convection simulations \citep{brun02,miesc06}, 
mean-field models \citep{kitch95,rempe05}, and
theoretical interpretation of helioseismic rotational 
inversions \citep{balbu09} suggest that the dominant
contributions to this equation are the Coriolis and
centrifugal terms associated with the differential 
rotation as well as the zonal component of the baroclinic 
vector.   The latter is proportional
to the mean latitudinal entropy gradient if the stratification
is nearly hydrostatic and adiabatic and if the equation of
state is that of an ideal gas.  Thus, neglecting the 
meridional components of the Reynolds stress, the Lorentz
force, the viscous diffusion, thermal fluctuations, and
quadratic terms in $\left<\vv_m\right>$ yields
\begin{equation}\label{eq:meridional}
\frac{\pd}{\pd t} \left<\omega_\phi\right> = \lambda \frac{\pd \Omega^2}{\pd z}
- \frac{g}{r C_P} \frac{\pd \left<S\right>}{\pd \theta} ~~~,
\end{equation}
where $\omega_\phi$ is the longitudinal component of the fluid vorticity 
$\vort = \curl \vv$, $S$ is the specific entropy, $g$ is the gravitational
acceleration, and $C_P$ is the specific heat at constant pressure.

We have retained the time dependence in eq.\ (\ref{eq:meridional}) in 
order to illustrate how mechanical and thermal forcing influence the 
evolution of the meridional flow.  If the balance in equation 
(\ref{eq:gp}) is not satisfied, this will lead to a change
in the specific angular momentum 
$\pd {\cal L}/\pd t = \lambda^2 \pd \Omega/\pd t$ that 
will in turn influence the meridional flow through the first
term on the RHS of eq.\ (\ref{eq:meridional}).  Likewise,
any imbalance in equation (\ref{eq:therm}) will change
the baroclinic (second) term through $\pd \left<S\right>/\pd t$.

The system will evolve nonlinearly toward thermal wind balance
(TWB), described by an equation that is now well known:
\begin{equation}\label{eq:twb}
\frac{\pd \Omega^2}{\pd z} = \frac{g}{r \lambda C_P} \frac{\pd \left<S\right>}{\pd \theta}  ~~.
\end{equation}
Most current theoretical and numerical models of solar mean flows 
attribute the conical nature of the solar differential rotation profile 
($\pd \Omega^2/\pd z \neq 0$) to latitudinal entropy gradients through 
eq.\ (\ref{eq:twb}).  In particular, \cite{balbu09} have argued that 
the orientation of the $\Omega$ isosurfaces in the solar CZ inferred from
helioseismology follows from thermal wind balance, Eq.\ (\ref{eq:twb}),
together with the hypothesis that isorotation and isentropic surfaces
coincide.  Furthermore, the existence of the near-surface shear layer
suggests that there is a transition near $r \sim 0.95 R$ below which
the Rossby number is less than unity and thermal wind balance prevails
(MH11).

Although we agree that it plays an essential role in determining the 
orientation of the angular velocity isosurfaces in the solar convection 
zone, we note that baroclinicity due to axisymmetric thermal gradients
cannot account for the existence of the solar differential rotation.  
The zonal component of the Reynolds stress must also contribute.  
This is demonstrated in the Appendix.

Here we wish to focus on the approach to thermal wind balance by means
of the time-dependent vorticity equation (\ref{eq:meridional}).
Although the meridional circulation itself, represented by $\left<\omega_\phi\right>$,
drops out of the balance equation in a steady state, any imbalance between <
the inertial and baroclinic terms on the RHS will induce a flow.  Consider
for example the northern hemisphere (NH).  There it is well known from
helioseimic observations that $\pd \Omega^2/\pd z < 0$
\citep{thomp03}.  This will tend to induce a counter-clockwise (CCW)
circulation ($\left<\omega_\phi\right> < 0$, poleward in the upper
convection zone) that will act to make the $\Omega$ profile more
cylindrical ($\pd \Omega / \pd z = 0$) in accordance with the
Taylor-Proudman theorem.  In order to oppose this and achieve TWB
[Eq.\ (\ref{eq:twb})], there must be a poleward entropy gradient ($\pd
\left<S\right>/\pd \theta < 0$ in the NH).  In short, the inertial
term tends to induce a CCW circulation in the northern hemisphere
while the baroclinic term tends to induce a clockwise (CW) circulation
(vice versa in the southern hemisphere).

In the approach to equilibrium, one of these terms must act as the driver,
accelerating the meridional flow, while the other acts as a resistance,
opposing the acceleration until a balance is achieved.  The observed
poleward sense of the meridional flow near the solar surface 
(\S\ref{sec:intro}) indicates that the driver is in fact the inertial 
term $\propto \pd \Omega/\pd z$.  As demonstrated in \S\ref{sec:gp},
this links the mean flows directly to the convective Reynolds stress 
and enables us to estimate the amplitude of the convective velocities
based on the observed differential rotation and meridional flow.

There is one potential caveat to this conclusion. The recent 
analysis by \citet{hatha11b} based on autocorrelation of 
photospheric Dopplergrams suggests that there may be a reversal
of the meridional flow at $r \sim 0.95 R$.  If this is indeed true, 
than we cannot rule out a CW circulation cell (NH, CCW in 
the south) in the deep convection zone driven by baroclinic forcing
(implying poleward flow near the base of the CZ).  However, this is 
in conflict with helioseismic inversions which suggest the meridional 
flow remains poleward well below 0.95$R$ 
\citep{giles97,braun98,chou01,beck02}.  Furthermore, poleward 
flow near the base of the CZ would be disasterous for 
flux-transport dynamo models (\S\ref{sec:ftd}).  This in 
itself does not preclude the presence of such a cell but
it demonstrates that poleward flow at the base of the 
convection zone is contrary to our current understanding 
of solar interior dynamics.

\section{Amplitude of Convective Velocities}\label{sec:amp}

\subsection{Fundamental Expressions}\label{sec:fund}

In this section we use Eq.\ (\ref{eq:bob}) to obtain lower limits 
on the amplitude of convective velocities throughout the solar convection 
zone.  The motions we are most interested in are the motions that are 
responsible for establishing the solar differential rotation by means of 
the Reynolds stress.  Thus, unlike solar granulation, they must be large
enough and slow enough to sense the rotational influence and spherical
geometry.  The mere existence of the solar differential rotation 
is evidence for the presence of such motions.  In what follows,
we will quantify what we mean by {\em large enough} and 
{\em slow enough}.

We begin by estimating the amplitude of the Reynolds stress inside the
divergence operator on the RHS of Eq.\ (\ref{eq:bob});
\begin{equation}\label{eq:rsamp}
\left<\rho \lambda \vv_m^\prime v_\phi^\prime\right> \sim \rho \lambda \epsilon V_c^2  ~~~.
\end{equation}
where $V_c$ is the characteristic amplitude of the convective velocity and 
$\epsilon = \left| \left<v_\phi^\prime\vv_m^\prime\right> \right| V_c^{-2}$
is a correlation coefficient describing the efficiency of the convective angular 
momentum transport.  If flows are perfectly correlated, the transport is
efficient and $\epsilon = 1$.  However, in practice $\epsilon$ will
fall somewhere between zero and one.  Convection simulations such as that
shown in Fig.\ \ref{fig:models} below yield $\epsilon \sim$ 0.1--0.2.
Similar values of are given by the mean-field theory of \cite{kitch05}.  
There the amplitude of the non-diffusive component of the Reynolds stress, 
$\Lambda$, is of order $\nu_t \Omega_0 {\cal H}$ where $\nu_t \sim V_c^2 \tau_c/3$ 
is the turbulent viscosity, $\tau_c$ is the correlation time, and ${\cal H}$ 
is a nondimensional normalization factor that depends on the inverse Rossby 
number $\Omega^* = 2 \tau_c \Omega_0$.  This yields 
$\epsilon \sim \Lambda V_c^{-2} \sim \Omega^* {\cal H}/6$, which according 
to their model, is about 0.1-0.2 in the lower CZ.

Note that we use a single velocity scale $V_c$ to characterize the
convection but this does not necessarily imply that the velocity field
is isotropic.  On the contrary, the velocity scale is defined in terms of the
Reynolds stress through equation (\ref{eq:rsamp}), which requires some 
degree of anisotropy.  Thus, $V_c$ can be regarded in general as the geometric 
mean of the two velocity components that dominate the Reynolds stress.

Throughout the paper we assume that the density $\rho$ is
spherically symmetric and is approximately given by Model S of
\citet{chris96}.  Then substituting (\ref{eq:rsamp}) into (\ref{eq:bob})
and noting that $0 \leq \epsilon \leq 1$ provides a lower limit on $V_c$ 
\begin{equation}\label{eq:vamp}
V_c \sim \left( \frac{\delta}{\epsilon} ~ V_m \left| \del {\cal L}\right| \right)^{1/2} 
\gtrsim \left(\delta ~ V_m \left| \del {\cal L}\right| \right)^{1/2} ~~~.
\end{equation}
where $V_m$ is a characteristic amplitude of the meridional flow 
and $\delta = L_t/\lambda$ where $L_t$ is the length scale associated
with the divergence operator in (\ref{eq:bob}) and $\lambda = r \sin\theta$
as above.

As above with the convection, we emphasize that the meridional flow is not isotropic.
This is particularly true in the upper and lower boundary layers of the CZ where the
meridional flow is predominantly horizontal and at the equator where it is predominantly
vertical.  The proper interpretation for $V_m$ is the amplitude of the flow across
${\cal L}$ isosurfaces, which are approximately cylindrical.  So, $V_m$ should be 
regarded as the flow toward or away from the rotation axis, which may be predominantly
radial or latitudinal depending on the latitude.

A useful variation of equation (\ref{eq:vamp}) can be obtained if we consider
only the component of ${\cal L}$ involving the mean rotation rate
$\Omega_0$.  This gives $\left| \del {\cal L}\right| \sim 2 V_\Omega$ 
where $V_\Omega = \lambda \Omega_0$ is the zonal velocity 
associated with the rotation of the star.  Substituting
this into Eq.\ (\ref{eq:vamp}) gives
\begin{equation}\label{eq:limit1}
V_c \gtrsim \left(2 \delta V_m V_\Omega\right)^{1/2}  ~~~~.
\end{equation}
Thus, if $(2 \delta)^{1/2}$ is of order unity, then
a lower limit on the convective velocity can be obtained
by taking the geometric mean between the meridional flow speed
and the rotational speed, relative to an inertial frame.
Taking nominal values of $V_m \sim $ 2--10 m s$^{-1}$,
$\Omega_0 = 2.7 \times 10^{-6}$ s$^{-1}$, and 
$\lambda \sim 0.85 ~ R / \sqrt{2}$ would then yield
$V_c \gtrsim$ 47--106 m s$^{-1}$.  However, as we will see
below, $(2 \delta)^{1/2}$ is likely to be somewhat less 
than unity, bringing these estimates down.
Still, the extension from (\ref{eq:vamp}) to (\ref{eq:limit1}) 
is robust, since the differential rotation serves to steepen
the ${\cal L}$ gradient relative to the uniform rotation 
value of $2 V_\Omega$ (see \S\ref{sec:estimates}).

It now remains to estimate $\delta$.  In this context, it is important
to emphasize that $L_t$ reflects the scale of turbulent transport, which
is not necessarily the same as the scale $L_c$ of the convective motions
themselves (the latter is addressed in \S\ref{sec:himp}).  If the
balance expressed by Eq.\ (\ref{eq:bob}) is to be realized, then
this scale must be intimately linked to the structure of the
meridional circulation.  Indeed, if our picture of gyroscopic 
pumping is correct then the structure of the meridional circulation
is largely {\em determined} by the spatial variation of the 
Reynolds stress, as reflected by $L_t$.

To illustrate this relationship, consider the simplest case of
a single meridional circulation cell in each hemisphere, with
poleward flow in the upper CZ and equatorward flow in the 
lower CZ.  Then the LHS of Eq.\ (\ref{eq:bob}) would
be correspondingly negative and positive in the upper
and lower CZ respectively.  To balance this, the RHS
must also change sign, suggesting $L_t \sim D/2$, where 
$D$ is the depth of the convection zone.  If ${\cal F}$
were to exhibit multiple sign changes across the CZ,
this would produce multiple layered circulation cells 
in radius such that $L_t \sim D/(2 N_c)$ where $N_c$ is
the number of cells.  For the moment arm we can consider
mid-latitudes in the mid-convection zone, so
$\lambda \sim r_m/\sqrt{2}$, where $r_m \sim 0.85 R$.  
Thus, we have
\begin{equation}\label{eq:delta}
\delta = \frac{\sqrt{2} L_t}{r_m} \sim \frac{D}{\sqrt{2} r_m N_c} \sim \frac{0.25}{N_c} 
\mbox{\hspace{.1in} ($V_m = \vert\left<v_\theta\right>\vert$ at 45$^\circ$ lat)}
\end{equation}
where we have used the value for the Sun of $D/R \sim 0.3$.  As noted, this
corresponds to poleward or equatorward flow at mid latitudes in one or more
circulation cells.  So, when used with Eq.\ (\ref{eq:vamp}) or 
(\ref{eq:limit1}) the value of $V_m$ should correspond to the latitudinal 
flow speed. 

The caveat that can potentially limit the applicability of eq.\ (\ref{eq:delta}) 
is that, although $N_c$ is regulated by the number of nodes in ${\cal F}$ 
(sign changes in the Reynolds stress divergence), the flow amplitude $V_m$
is regulated by local gradients.  This is particularly important at the base 
of the CZ where a strong convergence of the angular momentum flux over a narrow 
layer could drive an arbitrarily strong equatorward flow $V_m$ for a given $V_c$.  

\begin{figure}
\centerline{\epsfig{file=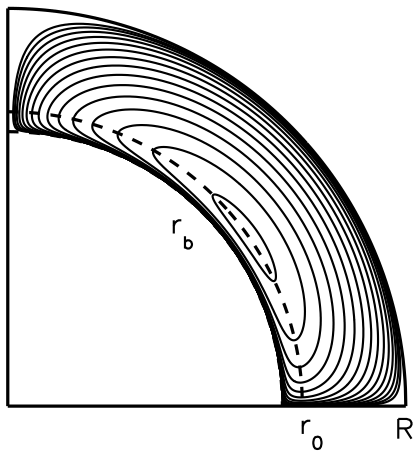,width=2.5in}}
\caption{Schematic diagram illustrating $r_0$ and $r_b$ 
for an idealized single-cell meridional circulation profile.
Both radii are indicated by dashed lines, although the latter is difficult
to distinguish from the streamlines of the mass flux, indicated by
solid lines.  The northern hemisphere is shown and the sense of the
circulation is counter-clockwise.  The turnaround radius $r_0$ marks 
a change in sign of the latitudinal flow while $r_b$ marks the 
radius below which the circulation is negligible.  The
radius of the Sun is denoted by $R$.  Thus, equatorward
flow occurs for $r_b < r < r_0$ and poleward flow for
$r_0 < r < R$.  \label{fig:mcfig}}
\end{figure}

To demonstrate this concept, consider a single-celled meridional circulation
profile ($N_c = 1$) as illustrated in Figure \ref{fig:mcfig}.  We define the
radius $r_0$ as the turnaround radius, where the poleward flow in the upper 
CZ transitions to the equatorward flow in the lower CZ.  We also postulate
the existence of a radius $r_b$ below which the amplitude of the meridional
circulation becomes negligible.  The existence of such a radius can be 
inferred based on the presence of the solar tachocline as deduced from 
helioseismology.  In the absence of other forces, the combined effects
of gyroscopic pumping, thermal wind balance, and radiative diffusion
would induce a meridional circulation below the convection zone that 
would wipe out the radial shear in the tachocline, causing the latitudinal
shear to spread downward on a time scale of several billion years 
\citep{spieg92}.  Helioseimic inversions indicate that this has not 
happened, suggesting the presence of some unknown torques that act 
to confine the tachocline \citep[e.g.][]{miesc05}.  Thus, the thinness
of the tachocline inferred from helioseismic rotation inversions suggests
that the amplitude of the meridional circulation likely drops off rapidly
below the base of the convection zone \citep{garau09}.  

So, in light of eq.\ (\ref{eq:gp}), $r_0$ separates the the region of 
poleward flow and flux divergence (${\cal F} < 0$), $r_0 \leq r \leq R$, 
from the region of equatorward flow and flux convergence
(${\cal F} > 0$), $r_b \leq r \leq r_0$.   Equation (\ref{eq:delta}) 
effectively assumes that the width of these layers is comparable, 
such that $r_0 - r_b \approx R - r_0$.  In other words, this equation
breaks down if the equatorward flow is confined to a much smaller 
region than the poleward flow, or vice versa.  In this case 
the effective $L_t = r_0 - r_b$ could become arbitrarily small,
implying small values of $\delta$ ($\ll 1$). A similar caveat also
holds for the multi-celled case $N_c > 1$. 

The mechanical forcing required to establish an asymmetric meridional
circulation profile such as that depicted in Figure \ref{fig:mcfig} 
via gyroscopic pumping may or may not be convective in origin
({\em asymmetric} in the sense that $r_0 - r_b << R - r_0$). 
A localized convergence of angular momentum flux could arise from 
the convective Reynolds stress as convective plumes are rapidly
decelerated by the subadiabatic stratification near the base
of the CZ.  Alternatively, it could arise from whatever 
non-convective processes may be involved in tachocline 
confinement, such as large-scale Lorentz forces or Reynolds
and Maxwell stresses induced by MHD instabilities 
\citep{spieg92,gough98,miesc05}.  Whatever its origin, 
the confinement mechanism would have to transport angular
momentum poleward in order to prevent the radiative spreading
of the tachocline.  This would imply a convergence of angular
momentum at mid to high latitudes that would in turn induce
a prograde torque and an equatorward flow by means of gyroscopic 
pumping.  This may plausibly be confined to a thin boundary layer 
in the vicinity of the tachocline.

Regardless of how thin the region of equatorward flow may be,
conservation of mass requires that there be a net poleward
flow in the remainder of the CZ, implying a flux divergence.
Might this also occur in a thin layer, such that $\delta << 1$?
If so, wouldn't equation (\ref{eq:vamp}) imply that $V_c$ could
be arbitrarily small?  In other words, could weak convection 
maintain the observed mean flows in the solar interior by
establishing a Reynolds stress that is nearly divergenceless
in the bulk of the CZ, with narrow regions of divergence
and convergence in the upper and lower boundary layers?  
In the remainder of this section we argue that the answer 
to this question is likely to be no.  Although this scenario 
is in principle consistent with the physics of gyroscopic pumping, 
it can be largely ruled out by helioseismic measurements.

We begin with a direct estimate of $\delta$ in the surface layers
obtained from helioseismic determinations of the meridional flow.  As
noted in \S\ref{sec:intro}, local heloiseismic inversions suggest that
the radius at which the poleward surface flow reverses sign is no
shallower than $0.95 R$, so $R-r_0 \gtrsim 0.05 R$.  If we equate this
with $L_t$, this provides a lower bound for $\delta$ in the surface
layers:
\begin{equation}\label{eq:delta2}
\delta \sim \sqrt{2} L_t / R \gtrsim 0.07 \mbox{\hspace{.2in} (surface layers)} ~~~.
\end{equation}
Thus, the factor of $\delta^{1/2}$ in eq.\ (\ref{eq:vamp}) is likely to 
be no smaller that 0.26 in the surface layers.

We now consider the nonlocal nature of gyroscopic pumping, which
provides a link between regions of Reynolds stress convergence and
divergence even if they are spatially separated (MH11).  Local forcing
can in principle induce a global meridional flow that couples the two
regions, even if they are localized in the upper and lower boundary layers.
Furthermore, since isosurfaces of ${\cal L}$ are approximately radial
at high latitudes, a single circulation cell with poleward flow at the
surface, downward flow near the poles, and equatorward flow at the
base of the CZ could in principle be sustained even if the
high-latitude Reynolds stress were divergenceless in the bulk of the
CZ (${\cal F} \approx 0$).
 
However, ${\cal F}$ cannot be zero in the bulk of the CZ
at low latitudes.  Mass conservation requires a radially
outward meridional flow that must be balanced by the
Reynolds stress.  If it were not, the circulation would 
quickly homogenize ${\cal L}$ at the equator, establishing an 
inward $\Omega$ gradient ($\Omega \propto r^{-2}$) on a 
crossing time scale $\sim D/V_m$ (about 6 years for 
$V_m \sim 1$ m s$^{-1}$).  Thus, the $\Omega$
profile and the horizontal divergence of the meridional
flow inferred from solar observations provide a robust 
diagnostic of the Reynolds stress at the equator.  We 
now exploit this diagnostic to obtain alternative estimates 
for $\delta$ and $V_c$.

Symmetry requires that the radial component of the flux 
divergence on the RHS of (\ref{eq:bob}) vanish at the 
equator, so $L_t$ must correspond to the latitudinal
convergence of the angular momentum flux that
sustains the radially outward $V_m$.  Since the convergence
of ${\cal F}$ must span all latitudes where the flow is
outward, we can estimate $L_t$ from the outward
mass flux $\dot{M} = \rho V_m 2 \pi r L_t$.  We have 
defined $\dot{M}$ to be the outward mass flux in
one hemisphere because in \S\ref{sec:estimates} we will 
equate it to the observed poleward mass flux in the surface 
layers.  Setting $\lambda = r$ at the equator then gives us an
alternate estimate for $\delta$
\begin{equation}\label{delta3}
\delta = \frac{L_t}{\lambda} = \frac{\dot{M}}{\rho V_m 2 \pi r^2} 
\mbox{\hspace{.1in} ($V_m = \vert\left<v_r\right>\vert$ at 0$^\circ$ lat)} ~~~.
\end{equation}
Substituting this expression into (\ref{eq:vamp}) yields
\begin{equation}\label{eq:limit2}
V_c \gtrsim \left( \frac{\dot{M} \left| \del {\cal L}\right|}{2\pi r^2 \rho }\right)^{1/2}  
\mbox{\hspace{.1in} ($V_m = \vert\left<v_r\right>\vert$ at 0$^\circ$ lat)}.
\end{equation}
Note that the meridional flow only appears in Eq.\ (\ref{eq:limit2}) through 
the mass flux $\dot{M}$, which we can estimate from helioseismic inversions
(\S\ref{sec:estimates}).    Since ${\cal L}$ and $\rho$ are also known from 
helioseismology and solar structure models, Eq.\ (\ref{eq:limit2}) provides 
a robust lower limit on the convective velocity near the equator that is
independent of $\delta$.  Given the observed uniformity of the solar 
irradiance \citep{rast08}, velocity amplitudes at higher latitudes are 
likely to be comparable.

\begin{figure}
\centerline{\epsfig{file=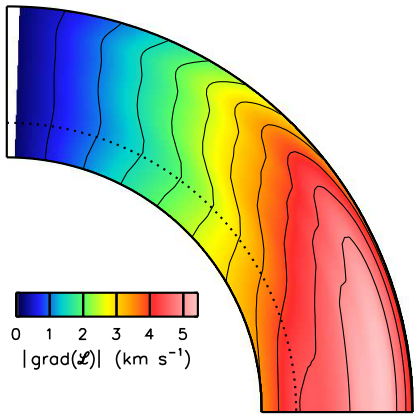,width=2.5in}}
\caption{Shown is the magnitude of the specific angular 
momentum gradient $\left| \del {\cal L}\right|$ inferred
from helioseismic rotational inversions.
For the corresponding profiles of $\Omega$ and ${\cal L}$
see Fig.\ 1 of MH11.  These results are based on RLS inversions
of GONG data from four non-overlapping intervals in 1996, 
provided by R.\ Howe \citep{howe00,schou02}.  The dotted 
line indicates the base of the convection zone.
\label{fig:Lfig}}
\end{figure}

In summary, Eq.\ (\ref{eq:limit2}) provides a reliable lower limit
on the convective velocity amplitude at low latitudes based firmly
in observable quantities.  The estimate is most reliable at a radius 
of $r \sim 0.95 R$ (above which we know that the meridional flow 
is poleward) but it can be extended to lower radii if a single-celled
profile ($N_c = 1$) is assumed.   Equation (\ref{eq:delta2}) is also
robust, depending only on helioseismic inversions in the surface layers.  
Together with Eqs.\ (\ref{eq:vamp}) and (\ref{eq:limit1}) it provides 
a reliable, though somewhat conservative, limit on the convective 
velocities above $0.95 R$.

Equation (\ref{eq:delta}) can be used together with Eqs.\ 
(\ref{eq:vamp}) and (\ref{eq:limit1}) to provide estimates
of $V_c$ in the lower CZ at mid-latitudes.  However, these estimates
break down if there is a large difference in the filling factor 
between polar and equatorial flows.  For example, if $N_c = 1$ and 
if the return equatorward flow near the base of the CZ is confined 
to a very small region such that $r_0-r_b \ll R - r_0$, then 
the effective $\delta$ near the base of the CZ could be significantly 
smaller than the estimate given in Eq.\ (\ref{eq:delta}).  This 
would in turn reduce the lower limit on $V_c$ near the base of the CZ 
according to Eqs.\ (\ref{eq:vamp}) and (\ref{eq:limit1}).

In the next section (\S\ref{sec:estimates}) we quantify our estimates 
by turning to photospheric observations and helioseismic inversions.
However, before proceeding, we briefly address the possibility of 
multiple circulation cells in latitude.

In short, the analysis and conclusions of this paper are insensitive
to the high-latitude structure of the meridional flow.  Our focus is
not on the mean flows themselves, but on using mean flow measurements
as tools to probe the convection.  Since helioseismic inversions of
differential rotation and meridional flow are most reliable at low to
mid latitudes, this is where we focus our attention.  These indicate a
single circulation cell near the surface up to latitudes of at least
50-60$^\circ$.  According to the gyroscopic pumping equation
(\ref{eq:gp}), the presence of one or more high-latitude counter-cells
may signify a change in the sense and possibly the amplitude of the
Reynolds stress, which may in turn signify a change in $V_c$.
However, given the small latitudinal variation of the solar irradiance
\citep{rast08}, it is more likely that a change in ${\cal F}$ at high
latitudes would be associated with a change in the efficiency factor
$\epsilon$, the transport scale $\delta$, the angular momentum
gradient $\vert \del {\cal L} \vert$, or other contributions 
to ${\cal F}$ such as Maxwell stresses.  These issues lie outside 
the scope of this paper.

\subsection{Estimates Based on Helioseismic Inversions}\label{sec:estimates}

In \S\ref{sec:fund} we gave a conservative estimate of 
$\left| \del {\cal L}\right|$ by considering only the uniform
rotation component.  A more precise estimate follows straightforwardly 
from helioseismic rotational inversions and is shown in 
Fig.\ \ref{fig:Lfig}.  At mid-latitudes its magnitude
is roughly 3 km s$^{-1}$, slightly larger than the value associated
with the uniform rotation component, 
$\left| \del {\cal L}_0\right| \sim2 V_\Omega \sim $ 2.7 km s$^{-1}$.
This is a consequence of the sense of the differential rotation,
prograde at the equator and retrograde at the poles, which enhances
the cylindrically outward $\del {\cal L}$.  Thus, the simple lower 
limit expressed in (\ref{eq:limit1}) is still valid for the Sun
when the differential rotation is taken into account.  The same
is true for any star with a solar-like differential rotation 
(equatorward $\del \Omega$).

An estimate for $V_m$ is harder to come by but we do have two 
pieces of reliable information from solar observations that
we will exploit.  First, the mean meridional flow at mid-latitudes is 
poleward on average near the surface for $r \gtrsim 0.95 R$, with 
an amplitude of about 10-20 m s$^{-1}$ (\S\ref{sec:intro}).  Second, 
solar structure models provide a reliable measure of the density throughout 
the convection zone that is verified by helioseismic structure inversions
to within a few percent \citep{christ02}.  We now proceed to use these 
two foundations together with mass conservation to obtain a lower bound 
on $V_m$.

We begin by estimating the poleward mass flux near the surface at
mid-latitudes
\begin{equation}\label{eq:Mdot}
\dot{M} = 2 \pi \int_{r_s}^R \left<\rho v_\theta\right> r dr \equiv
2 \pi ~ \widetilde{V}_\theta ~ \int_{r_s}^R \rho r dr ~~~,
\end{equation}
where $r_s = 0.95 R$ and we will take $\rho$ from Model S 
(see \S\ref{sec:fund}).  In order to obtain a lower limit on $V_m$, and thus, the 
convective velocity $V_c$, we take the mass-weighted poleward flow near the surface 
to be $\widetilde{V}_\theta \sim $ 10 m s$^{-1}$, which gives 
$\dot{M} \sim 4.1\times 10^{21} \mbox{g s$^{-1}$}$.

We treat $\dot{M}$ as a known quantity and estimate the average (density-weighted) 
deep return flow $V_m$ based on mass conservation as
\begin{equation}\label{eq:Vd}
V_m \sim \dot{M}
\left(2 \pi \int_{r_b}^{r_0} \rho r dr \right)^{-1}  ~~~.
\end{equation}
Here $r_b$ is the radius below which the meridional
flow is negligible and $r_0$ is the turnaround radius where the mean flow
shifts from poleward to equatorward (see Fig.\ \ref{fig:mcfig} and the
associated discussion in \S\ref{sec:fund}).  Note that this does not preclude 
multiple cells in radius; regardless of the meridional flow profile, 
there must be a net equatorward flow below $r_s$ to balance the poleward
flow above $r_s$.  Note also that some component of this return flow
must necessarily cross ${\cal L}$ contours (which are approximately
cylindrical: see MH11), requiring a Reynolds stress to maintain it 
as expressed in Eq.\ (\ref{eq:bob}).

\begin{figure}
\centerline{\epsfig{file=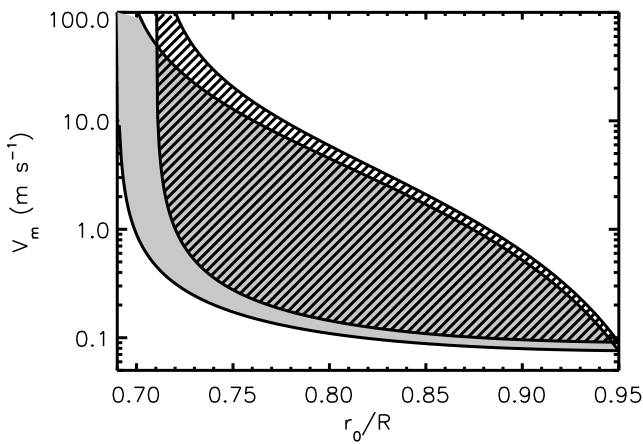,width=3.5in}}
\caption{Magnitude of the deep meridional flow $\widetilde{V}_m$ as a 
function of the turnaround radius $r_0/R$ estimated from 
Eq.\ (\ref{eq:Vd}) for $r_b = 0.69 R$ (grey shaded area) 
and $r_b = 0.71$ (hatched area).  Each area is bounded by the lower curve,
given by $\dot{M} = 4.1 \times 10^{21}$ g s$^{-1}$ and an upper curve where
$\dot{M}$ is given by Eq.\ (\ref{eq:Mdot}) with $r_s$ replaced by $r_0$
and $\widetilde{V}_\theta = $ 10 m s$^{-1}$.\label{fig:vtheta}}
\end{figure}

Figure \ref{fig:vtheta} shows estimates of $V_m$ based on two different
assumptions for the base of the main circulation cell, $r_b$.  A robust 
lower limit for $V_m$ is obtained by setting $\dot{M}$ to be 
the value known for the near-surface flow above $0.95 R$.
An upper limit follows if we assume that the 10 m s$^{-1}$ 
poleward flow persists for the entire region above the
turnaround radius $r > r_0$.  Then the deep flow $V_m$ must
be correspondingly stronger to balance the greater poleward
mass flux.  The actual meridional flow speed is likely to lie 
between these limits, as indicated in Fig.\ \ref{fig:vtheta}.

We are now able to obtain quantitative limits on $V_c$ based on 
the expressions derived in \S\ref{sec:fund}.  
Taking $\left| \del {\cal L}\right| \sim $ 3 km s$^{-1}$ 
at mid latitudes from Fig.\ \ref{fig:Lfig} and $\delta$ from 
Eq.\ (\ref{eq:delta}) and substitute them into Eq.\ (\ref{eq:vamp})
gives
\begin{equation}\label{eq:qlimit1}
V_c \gtrsim 27 ~ \mbox{m s$^{-1}$} ~ N_c^{-1/2} \left(\frac{V_m}{\mbox{1 m s$^{-1}$}}\right)^{1/2} ~~.
\end{equation}
Thus, if the latitudinal flow speed $V_m$ lies somewhere between 
0.1--10 m s$^{-1}$ as suggested by Figure \ref{fig:vtheta}, then 
the lower limit for $V_c$ in Eq.\ (\ref{eq:qlimit1}) lies somewhere 
between 8.6 -- 86 m s$^{-1}$.

The higher end of the range, $\sim $ 86 m s$^{-1}$, will apply near 
the surface ($r \gtrsim 0.95R$) where the meridional flow speed is 
known to be 10--15 m s$^{-1}$.  The lower estimate of 8.6 m s$^{-1}$ 
only applies if the meridional flow speed is as slow as 0.1 m s$^{-1}$ 
near the base of the CZ, if it penetrates deeper than $0.69 R$, or 
if the circulation profile is multi-celled ($N_c > 1$).  Lower
values of $V_c$ are also possible if the meridional flow near the 
base of the convection zone is confined to a narrow layer in radius
as described in \S\ref{sec:fund}.  Near the top of the CZ, 
Eqs.\ (\ref{eq:delta2}) and (\ref{eq:vamp}) give a more
reliable but more conservative estimate of 
$V_c \gtrsim $ 45 m s$^{-1}$.

Recall that these estimates are based on the latitudinal flow
speed at mid latitudes.  By contrast, the estimate given in 
Eq.\ (\ref{eq:limit2}) is based on the radial flow
at the equator.  Thus, for this estimate, we use the
larger equatorial value of 
$\left| \del {\cal L}\right| \sim $ 5 km s$^{-1}$ 
indicated by Fig.\ \ref{fig:Lfig}.  We also equate the
upward mass flux at the equator with the poleward mass flux
above $r \sim 0.95 R$ inferred from local helioseismology.
As noted above, this gives a value of 
$\dot{M} = 4.1 \times 10^{21}$ g s$^{-1}$.  Substituting
these values into Eq.\ (\ref{eq:limit2}) yields
\begin{equation}\label{eq:qlimit2}
V_c \gtrsim 30 ~ \mbox{m s$^{-1}$} ~ 
\left[\frac{\rho}{0.008 ~ \mbox{g cm$^{-3}$}}\right]^{-1/2}
\left[\frac{r}{0.95 R}\right]^{-1}
\end{equation}
As noted in \S\ref{sec:fund}, this estimate is most reliable
at $r \sim 0.95 R$ above which the meridional flow is known
to be poleward so we have a reliable estimate for $\dot{M}$.
The value of $\rho = 0.008$ g cm$^{-3}$ used in Eq.\ 
(\ref{eq:qlimit2}) is based on Model S at $r \sim 0.95 R$.
If we assume this outward mass flux extends down to the
lower CZ, then Eq.\ (\ref{eq:qlimit2}) implies 
$V_c \gtrsim $ 9 m s$^{-1}$ at $r \sim 0.75 R$ (where
$\rho \sim 0.14$ g s$^{-1}$).

Combining these seperate estimates, we conclude that
the amplitude of convective velocities must be at 
least 30 m s$^{-1}$ in the upper CZ ($r \sim 0.95 R$)
and at least 8 m s$^{-1}$ in the lower CZ ($r \sim 0.75 R$).
These of course are lower limits so larger amplitudes 
are entirely possible and even likely.

\section{Discussion}\label{sec:discussion}

\subsection{Length Scales and Implications for Detectability}\label{sec:himp}

The lower limits obtained for convective velocities in \S\ref{sec:amp} make no
reference to the characteristic length scale at which the convective motions
occur.  In this section we address this issue and consider its implications
for the detectability of deep convection by means of local helioseismology.

Some insight into the characteristic length scale $L_c$ for deep convection 
can be obtained by considering the Rossby number
$R_o = V_c/(2 \Omega L_c)$. As is well known, $R_o$ quantifies the amplitude 
of advective momentum transport in a rotating reference frame relative
to the Coriolis force; small values $R_o \lesssim 1$ imply strong rotational 
influence and large values weak.

Numerical simulations of convection exhibit a profound change in the
nature of convective angular momentum transport as the Rossby number
varies across unity.  When the rotational influence is weak $R_o > 1$, 
convective flows tend to conserve their angular momentum locally, 
producing anti-solar differential rotation profiles in which the angular 
velocity increases toward the rotation axis, implying $\pd \Omega/\pd \theta < 0$
in the NH
\citep{gilma77,gilma79b,hatha82,deros02,aurno07,augus11,feath12}.  As the 
rotational influence becomes stronger, the preferred convective modes tend 
to align with the rotation axis, and the resulting velocity correlations induce
an equatorward angular momentum transport by means of the convective Reynolds
stress \citep{busse02,miesc09}.  This generally promotes solar-like angular 
velocity profiles in which the angular velocity gradient is equatorward
($\pd \Omega/\pd \theta > 0$ in the northern hemisphere).  It is 
currently an open question precisely where this transition lies but
simulations suggest that it occurs for values of $R_o$ somewhat
less than unity, perhaps around 0.6 $\pm 0.3$ \citep{feath12}.

Although this insight is based on convection simulations (which, like 
any model, have limitations), there is good reason to suspect that it 
still applies in the extreme parameter regime of the solar interior.
Consider a radial downflow convective plume that is
accelerated in the upper thermal boundary layer of the Sun, near the
photosphere.  The Coriolis force operating on this plume will
tend to deflect it in a prograde direction with a characteristic
time scale of $\tau \sim (2 \Omega_0)^{-1}$ and a radius of 
curvature $r_c = V_c/(2 \Omega_0)$.  If the vertical coherence of the 
plume in the absence of rotation is $L_c$ then we obtain 
$r_c/L_c = R_o$.

To appreciate the significance of this, consider the equatorial plane
($\sin\theta = 1$) and set $L_c$ to be the depth of the convection zone.
If $R_o \gg 1$, the plume is deflected only slightly in a prograde
direction before it reaches the base of the convection zone.  This
will induce a negative $\left<v_r^\prime v_\phi^\prime\right>$
correlation, producing cylindrically inward angular momentum transport
and an anti-solar differential rotation profile.  However, if $r_c$ 
is less than the depth of the CZ, then our idealized, ballistic
plume will never make it 
to the bottom.  The vertical dominance of the flow will be broken and 
the nature of the Reynolds stress will be profoundly altered.  This 
argument can be readily generalized to any intrinsic vertical coherence 
length $L_c$ and any latitude, with an effective radius of curvature
of $r_c/\sin\theta$.

More generally, it is reasonable to argue that only motions
with $R_o < 1$ will possess a rotational influence strong enough
to establish the solar differential rotation (for $r \lesssim 0.95 R$).
The corresponding 
length scale $L_c \sim V_c / (2 \Omega_0 R_o)$ can be expressed in terms
of the spherical harmonic degree $\ell \sim 2 \pi r / L_c$, 
as follows;
\begin{equation}\label{eq:ell}
\ell \sim \frac{4\pi r \Omega_0}{V_c} ~ R_o 
\lesssim 750 ~ R_o ~ \left[\frac{V_c}{\mbox{30 m s$^{-1}$}}\right]^{-1} ~~. 
\end{equation}
For the numerical estimate we have used $\Omega_0 = 2.7 \times 10^{-6}$
and $r = 0.95 R$.  Note that $\ell = 750$ corresponds to a physical
length scale of 5.5 Mm.

Thus, setting $R_o < 1$ in Eq.\ (\ref{eq:ell}) suggests that the
convective motions responsible for maintaining the solar differential
rotation and meridional circulation must occur at spherical 
harmonic degree no greater than 750, implying a length scale
no less than 5.5 Mm.  

Note that the values used to obtain the numerical estimate in 
Eq.\ (\ref{eq:ell}), namely $r = 0.95 R$ and $V_c \gtrsim 30$ m s$^{-1}$ 
correspond to the base of the near-surface shear layer where the
transition from large to small Rossby number apparently occurs
(MH11). Note also that these values are almost certainly a 
conservative estimate.  If the transitional Rossby number is
closer to 0.5, the convective velocity scale is closer to 
50 m s$^{-1}$ and the efficiency factor in Eq.\ (\ref{eq:rsamp}) 
is closer to 0.5, then the lower limit in Eq.\ (\ref{eq:ell}) 
assumes plausible values of $\ell \sim 160$ and 
$L_c \sim $ 25 Mm.  One might also argue that 
convective motions below $0.95 R$ are likely to occupy scales
no smaller than supergranules, which would imply $\ell \lesssim 130$ 
and  $L_c \gtrsim $ 30 Mm.

These latter values are remarkably close to the the density and 
pressure scale heights, $H_\rho$ and $H_P$, which provide an 
independent estimate for $L_c$ by virtue of mixing length theory.  
According to Model S, $H_\rho \sim $ 20 Mm and $H_P \sim $ 12 Mm
at $r = 0.95 R$, increasing steadily to 90 and 57 Mm respectively 
at the CZ base.  Interestingly, if we assume that the length
scale of convective motions at $r = 0.95 R$ is equal to the 
density scale height, $L_c \sim $ 20 Mm, then this implies that the
transition from strong to weak rotational influence that marks
the base of the NSSL occurs at a Rossby number of about 0.28.

More generally, mixing-length theory predicts that $L_c$ should increase 
with decreasing $r$, becoming larger near the base of the CZ. A naive
application of Eq.\ (\ref{eq:ell}) suggests the opposite; inserting
our estimate of $V_c \gtrsim $ 8 m s$^{-1}$ in the lower CZ would
push the limit toward higher $\ell$.  However, this is a 
misapplication of Eq.\ (\ref{eq:ell}).  Convection simulations
and mixing length theory suggest that the Rossby number should
decrease toward the base of the CZ more rapidly than $V_c$,
implying an increase in $L_c$.

As mentioned in \S\ref{sec:intro}, recent work by \cite{hanas10,hanas12}
based on local helioseismic inversions suggests that the spectral
amplitude of convective motions may be no more than 1 m s$^{-1}$ 
on scales $\ell \lesssim 60$ at a radius of $r \sim$ 0.92--0.95 $R$.  
Smaller horizontal scales (higher $\ell$) lie beyond their detection 
limits at that depth.  This is difficult to reconcile with our lower 
limit of 30 m s$^{-1}$ but may be possible if deep solar convection
occupies multiple disparate scales, with broad, weak upflows surrounding
very narrow downflows.  This may well yield small spectral amplitudes
for global-scale modes ($\ell < 60$) while local velocities in downflows
might be much higher.  However, this still poses significant challenges
to our current paradigm for how solar mean flows are maintained.

In summary, the lower limits on convective velocities obtained
in \S\ref{sec:amp} based on the maintenence of mean flows and
the upper limits obtained by \cite{hanas10} based on local 
helioseismic inversions are both consistent with the idea that 
the characteristic length scale of convective motions is 
comparable to the local scale heights $H_\rho$ and $H_P$.  
This yields Rossby numbers less than unity in the deep 
convection zone where mean flows are established
and Rossby numbers greater than unity in the near-surface 
shear layer, as suggested by solar observations (MH11).

Finally, we note that the estimated length scales
discussed here bode well for the future of global
solar convection simulations. 
In order to adequately resolve a structure of 
size $L_c$, a simulation should have a resolution of at 
least $\sim 0.1 L_c$.  Thus, if $R_o \sim 1$, 
Eq.\ (\ref{eq:ell}) suggests that a resolution as 
high as $\ell \sim 7500$ (grid spacing $\delta \sim$ 0.65 Mm) 
could be required to capture the relevant dynamics.  
This is beyond the current capabilities of global models 
but recall that this is a conservative limit.  For plausible 
values in the mid CZ of $R_o \sim 0.2$, $V_c \sim $ 50 m s$^{-1}$ 
and $\epsilon \sim 0.5$, Eq.\ (\ref{eq:ell}) implies 
$\ell \lesssim 64$ ($L_c \gtrsim $ 65 Mm).  Thus,
a resolution extending to $\ell \sim 640$ should 
be sufficient to capture the dominant physical scales.
This is achievable now with current global simulations 
\citep{miesc08}.

This is not to say that current simulations necessarily 
capture all of the relevant dynamics.  But, to put 
it colloquially, they're beginning to approach the right 
ballpark. This suggests that we may be near a threshold in the
sense that moderate increases in resolution, together 
with improved modeling of the surface boundary layer
and the overshoot region, could yield significant 
advances in our understanding of solar convection
and the mean flows it establishes.  This may be achieved
both through the availability of increasingly powerful
computing resources and through improved numerical algorithms
that achieve higher parallel efficiency.

\subsection{Consistency with Convection Models}\label{sec:models}

We emphasize again that the limits on the convective velocities 
deduced in \S\ref{sec:fund} and \S\ref{sec:estimates} do not depend 
on any theoretical or numerical model other than the basic MHD equations 
and the dynamical balance expressed by Eq.\ (\ref{eq:bob}).  
Thus, it is of interest to ask whether theoretical and numerical 
models of convection are indeed consistent with these limits.

\begin{figure}
\centerline{\epsfig{file=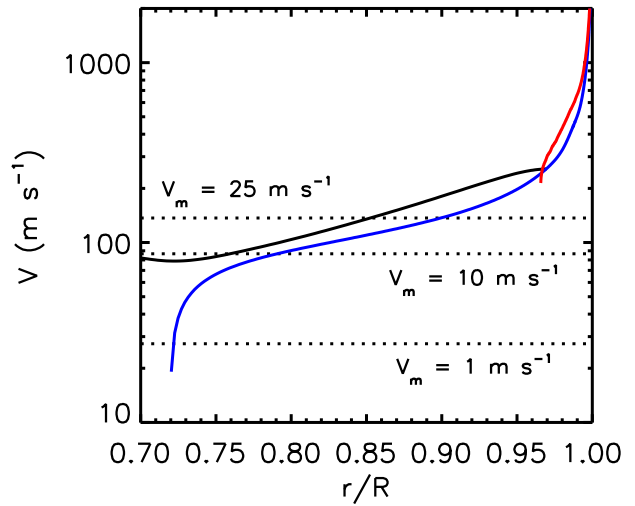,width=3.5in}}
\caption{Comparison of theoretical lower limits for the convection
amplitude (dotted lines) with numerical and theoretical models
of convection (solid lines).  The three dotted lines are obtained from 
Eq.\ (\ref{eq:qlimit1}) with $V_m = $ 1.0, 10 and 25 
m s$^{-1}$ as indicated.  The black line is obtained
from a global convection simulation with the ASH code 
\citep[described in][]{weber11} and the brown line is from a simulation
of surface convection done with the MURAM code \citep{vogle05,rempe09}. 
The blue line is from a composite model combining surface convection
simulations with the STAGGER code ($r > 0.97R$) with a mixing length
model ($r < 0.97R$), the latter calibrated to give the same entropy
jump as the simulation and scaled for continuity 
\citep{tramp11}.\label{fig:models}}
\end{figure}

The short answer is yes; models of solar convection based on 
several disparate physical perspectives and modeling approaches 
are uniformly consistent with the ideas propoposed in this paper.  
This is demonstrated in Figure \ref{fig:models}.

Shown in the Figure are results from global convection simulations
based on the ASH code (black line, see Weber et al.\ 2011 for more 
information about this particular simulation), surface convection 
simulations based on the MURAM code (red line, see 
V\"ogler et al.\ 2005 and Rempel et al.\ 2009 for a description 
of the model and Rempel 2011 for more details on this series of
simulations), and a hybrid model that combines surface convection 
simulations (from the STAGGER code) with a deep extrapolation based 
on mixing-length theory (blue line, see Trampedach \& Stein 2011 
for more information).  Note that both surface convection simulations 
presented here correspond to the quiet sun, with no active regions or 
flux emergence.\nocite{rempe11}

Superposed on the various curves are horizontal lines representing 
the theoretical lower limit for $V_c$ obtained from Eq.\ 
(\ref{eq:qlimit1}) with $V_m$ = 1.0, 10, and 25 m s$^{-1}$.  
All curves lie above the first limit, implying that the 
convective motions are at least in principle strong enough to 
sustain a solar-like differential rotation with a meridional 
flow of order 1 m s$^{-1}$ or less in the deep CZ.  
However, they are only strong enough to sustain a meridional 
flow speed $\gtrsim$ 25 m s$^{-1}$ in the upper half of the
CZ. So, the convective amplitudes are consistent with mean 
flows inferred from solar observations and they confirm the 
expectation from mass conservation that the flow speed of the 
deep equatorward return flow is likely to be less than the 
poleward surface value of 10--20 m s$^{-1}$.  Furthermore, 
all curves satisfy the limits obtained from 
Eq.\ (\ref{eq:qlimit2}) of $V_c \gtrsim $ 30 m s$^{-1}$ and 
$V_c \gtrsim $ 9 m s$^{-1}$ in the upper and lower CZ respectively
and all curves satisfy the limit of $V_c \gtrsim $ 45 m s$^{-1}$ 
for $r \gtrsim 0.95 R$ obtained from eqs.\ (\ref{eq:vamp})
and (\ref{eq:delta2}).

Further confirmation of the consistency of these models comes
from mean-field models of the solar differential rotation
and meridional circulation by \cite{rempe05,rempe06}.
Here the convective amplitudes are essentially prescribed
{\em a priori} by means of the imposed Reynold stress, modeled 
as a turbulent diffusion plus a $\Lambda$-effect.  This is
in contrast to the convection simulations shown in 
Figure \ref{fig:models} where $V_c$ is a product of the 
simulation.  However, these mean-field models can be calibrated 
to produce solar-like mean flows so estimates of $V_c$ can be 
obtained by selecting the optimal transport coefficients.
Here we are interested in particular in the $\Lambda$-effect, 
which is the non-diffusive component of the Reynolds stress
tensor responsible for maintaining the differential rotation
and, ultimately, the meridional circulation by means of
gyroscopic pumping (\S\ref{sec:gp}).

In Rempel's models, the amplitude of the $\Lambda$-effect
is given by $\Lambda_0 \nu_t \Omega_0$ where $\Lambda_0$ is 
a non-dimensional coefficient of order unity and $\nu_t$ is 
the turbulent viscosity.  Relating this to the convective 
velocity as in Eq.\ (\ref{eq:rsamp}) yields 
$V_c \gtrsim (\Lambda_0 \nu_t \Omega_0)^{1/2}$ for $\epsilon \leq 1$.
Solar-like mean flows are generally obtained with
$\Lambda_0 \sim 0.8$ and 
$\nu_t \sim$ 3--5 $\times 10^{12}$ cm$^2$ s$^{-1}$
in the bulk of the CZ.  Near the base of the CZ,
$\nu_t$ drops sharply by more than an order of magnitude.

The value of $\nu_t \sim$ 3--5 $\times 10^{12}$ cm$^2$ s$^{-1}$
in the upper CZ implies $V_c \gtrsim $ 25--35 m s$^{-1}$.
This is comparable to our estimate of $V_c \gtrsim $ 
30 m s$^{-1}$ based on Eq.\ (\ref{eq:limit2}).  However,
this lower limit lies below the limit of 
$V_c \gtrsim $ 86 m s$^{-1}$ in the upper CZ based on 
Eq.\ (\ref{eq:qlimit1}) and $V_c \gtrsim $ 45 m s$^{-1}$ 
based on Eqs.\ (\ref{eq:vamp}) and (\ref{eq:delta2}).
This implies that the effective value of $\delta$ in the
upper CZ is somewhat smaller than that given by Eqs.\ 
(\ref{eq:delta}) and (\ref{eq:delta2}).  Since $\nu_t$ is
nearly constant in the upper CZ, this smaller value of 
$\delta$ can be attributed to the density gradient,
which factors into the Reynolds stress as indicated
in Eq.\ (\ref{eq:bob}).  Similarly, a strong convergence 
of the angular momentum flux ($\delta \ll 1$) near the 
base of the CZ (associated with the sharp drop in $\nu_t$) 
sustains an equatorward flow of a few m s$^{-1}$ despite 
the low value of $\nu_t$.

The value of $\nu_t$ used by Rempel was deliberately chosen
to be relatively small in order to make it more compatable
with the small value of the turbulent magnetic diffusivity
$\eta_t \sim 10^{11}$ cm$^2$ s$^{-1}$ in the mid CZ, needed
for the operation of the advection-dominated flux-transport
dynamo (\S\ref{sec:ftd}).  In many mean-field theories such 
as that of \cite{kitch05}, $\nu_t$ is significantly larger, 
often exceeding 10$^{13}$ cm$^2$ s$^{-1}$ as predicted by
mixing-length theory.

The consistency among these disparate models is remarkable.  
The surface convection simulations (blue and red curves in
Fig.\ \ref{fig:models}) make good contact with photospheric 
observations of convective amplitudes but they do not address
mean flows; differential rotation 
and meridional circulation lie outside the scope of the models.
Meanwhile, the global convection simulations self-consistently
produce a solar-like differential rotation and the convective
velocity amplitudes roughly match the surface convection
simulations (to within about 20\%) in their small region 
of overlap near $r \sim 0.97 R$.  Furthermore, the global
(ASH) simulation satisfies the constraint in Eq.\
(\ref{eq:vamp}) with $\delta \sim 0.25$ and meridional flow 
speeds are consistent with Figure \ref{fig:vtheta}, with values 
of a few m s$^{-1}$ near the base of the CZ and 
$\sim$ 10--20 m s$^{-1}$ near the surface.  The mean-field 
models by \cite{rempe05,rempe06} produce solar-like mean
flows with Reynolds stress amplitudes somewhat lower than
suggested by the convective models but they are consistent
with an efficiency factor of $\epsilon < 1$.  Furthermore,
they are still consistent with the velocity estimates put 
forth in $\S\ref{sec:amp}$ when one takes into account the
relatively sharp gradients in the imposed angular momentum 
flux (small $\delta$), particularly at the base of the CZ.

We also note that our lower limit for $V_c$ of 
8 m s$^{-1}$ at the base of the CZ is consistent
with the upper limit of $V_c < 50$ m s$^{-1}$ obtained
by \citet{isik09} based on numerical models of the 
interaction between thin flux tubes and convective flows.
Larger convection amplitudes disrupt the flux storage,
promoting buoyancy instabilities on time scales shorter
than the dynamo amplification time of a few years.  

\subsection{Implications for Flux-Transport Dynamo Models}\label{sec:ftd}

Arguably one of the most successful (and certainly one of the most popular) 
current paradigms for modeling the origin of the solar activity cycle is 
the flux-transport dynamo.  For recent reviews see \cite{dikpa09} and
\cite{charb10} and for further details see 
\cite{wang91b,choud95,dikpa99b,dikpa01c,dikpa06,kuker01,nandy01,bonan02,rempe06,jouve07,jiang07,yeate08,guerr09,munoz09,munoz11,hotta10},
and many more.

Flux-transport (FT) dynamo models are mean-field models that solve the axisymmetric
MHD induction equation, typically in the kinematic limit.  Like other mean-field 
models, they involve specified prescriptions for magnetic field generation and 
transport by non-axisymmetric motions such as convection and flux emergence that 
lie outside the scope of the model.  Their defining characteristic that sets
them apart from other mean-field models is that the mean meridional circulation
plays an essential role in transporting magnetic flux and in thereby regulating
the cycle period.

Of particular importance is the direction and speed of the meridional flow
near the base of the convection zone.  There, an equatorward flow with 
a speed of 2-4 m s$^{-1}$ provides a robust mechanism for producing solar-like
butterfly diagrams, whereby the mean toroidal field (taken as a proxy for
sunspots) migrates toward the equator on a time scale of about 11 years.

Most FT models are also Babcock-Leighton models in which the emergence
and subsequent dispersal of active region flux in the photosphere acts
as a source for mean poloidal field \citep[e.g.][]{charb10}.
If this is the principal source of poloidal flux and if the mean
toroidal field is generated near the base of the convection zone as in
most models, then the dynamo requires a transport mechanism in order
to operate.  The two principal transport mechanisms considered in the
literature are the meridional circulation and turbulent transport, the
latter typically represented as a turbulent diffusion or magnetic
pumping.  Thus, Babcock-Leighton FT models may be further classified
as advection-dominated or diffusion-dominated depending on which of
these transport mechanisms plays a larger role \citep{yeate08,dikpa09}.

In short, the meridional circulation in advection-dominated FT dynamo
models serves two roles.  First, it regulates the cycle period
through the equatorward transport of toroidal flux near the base of
the CZ and second, it couples the spatially separated source regions
for mean poloidal and toroidal flux.  These two roles are distinct but
they are not independent.  For example, if the transport mechanism
coupling poloidal and toroidal sources is too efficient, it can
``short-circuit'' the dynamo, reducing the cycle period by decreasing
the effective path length at the base of the CZ over which the equatorward 
advection of toroidal flux operates.

The advection-dominated regime generally works well in the sense that 
it compares well with solar observations.  Imposing a single-celled
meridional flow in radius ($N_c = 1$) with a poleward flow speed
of 10-20 m s$^{-1}$ near the surface as indicated by observations
(\S\ref{sec:intro}) and an equatorward return flow of 2-4 m s$^{-1}$
near the base of the convection zone (consistent with 
Fig.\ \ref{fig:vtheta}) generally produces solar-like magnetic
cycles with a duration of about 11 years when used in conjuction
with a solar-like differential rotation profile and a 
Babcock-Leighton source.  More subtle issues such as dynamo parity,
saturation, and cycle modulation have generally been handled through 
minor variations on this basic paradigm.  For example, dipolar parity 
can be promoted by an additional source of poloidal flux deeper in 
the CZ \citep{dikpa01c} or by enhanced turbulent diffusion in the
surface layers \citep{hotta10}.

Although their empirical success and simplicity is rather
compelling, the principle problem with FT dynamo
models has always been the theoretical justification of the 
advection-dominated regime.  In order for the meridional circulation
to dominate the flux transport, convective transport must be relatively 
inefficient.  Advection-dominated FT models typically model the
latter as a turbulent diffusion and set the amplitude of the turbulent
diffusivity $\eta_t$ in the mid convection zone to be 
$\sim$ 10$^{11}$ cm s$^{-2}$ or less.  This is at least two 
orders of magnitude smaller than estimates of $\eta_t$ based
on mixing-length theory and convection simulations; see
\cite{munoz11} for a more detailed discussion of the 
problem.

Our estimates of the convective velocity and length scales in
\S\ref{sec:amp} provide a measure of the convective transport that is
independent of mixing length theory and convection simulations.
Moreover, these estimates are directly linked to the amplitude
and profile of the differential rotation and meridional circulation, 
which are essential ingredients in all FT dynamo models.  Thus, to 
explore the implications of these estimates, we can take the actual 
mean flow profiles used in FT models as a starting point and then 
ask how strong the convective motions must be in order to maintain 
these flows.

The essence of the problem can be appreciated merely from
equation (\ref{eq:limit1}).  This suggests that the convective
velocity must be of order the geometric mean between the rotational
velocity and the meridional flow velocity, provided that the 
transport scale $\delta$ is not inordinately small.  As noted
there, this implies $V_c \gtrsim$ 47--106 m s$^{-1}$ for the
meridional flow speeds of 2--10 m s$^{-1}$ typically used in 
FT dynamo models. If we then assume that the size scale of the 
convection $L_c$ is of order the density scale height of
60 Mm in the mid CZ, we obtain 
$\eta_t \sim V_c L_c/3 \sim 4\times 10^{11}$--$2 \times 10^{12}$ 
cm$^2$ s$^{-1}$.  This is significantly higher than the values
used in the lower CZ for many FT dynamo models.   In what follows
we provide a more detailed exposition of this result and its
implications.

Thus, we begin with a standard FT dynamo model with a single-celled 
profile ($N_c = 1$) and a solar-like differential rotation.  If the
turnaround radius $r_0$ is near the middle of the CZ, then $\delta$
is given by Eq.\ (\ref{eq:delta}) and a lower limit on $V_c$ is 
given by Eq.\ (\ref{eq:qlimit1}).  Setting $V_m \sim $ 2 m s$^{-1}$ 
near the base of the CZ implies $V_c \gtrsim $ 38 m s$^{-1}$.
In the upper CZ we can set $V_m \sim $ 10 m s$^{-1}$ 
which implies $V_c \gtrsim $ 86 m s$^{-1}$.

By combining these estimates of the velocity scale with the
limits on the length scale $L_c$ discussed in \S\ref{sec:himp},
we can obtain an estimate of the turbulent magnetic 
diffusivity $\eta_t \sim V_c L_c  / 3$.  Taking the very conservative
limit of $L_c \gtrsim $ 5.5 Mm gives 
$\eta_t \gtrsim 7 \times 10^{11}$ cm$^2$ s$^{-1}$ near 
the bottom of the CZ and 
$\eta_t \gtrsim 10^{12}$ cm$^2$ s$^{-1}$ near 
the top.  Although this is somewhat smaller than estimates based 
on mixing-length theory \citep[$\sim 10^{13}$, see][]{munoz11}, 
it is still nearly an order of magnitude larger than the 
values typically used in advection-dominated FT dynamo models.  
Using an estimate for $L_c$ based on density or pressure scale 
heights increases $\eta_t$ by nearly another order of magnitude
(\S\ref{sec:himp}).

As noted in \S\ref{sec:fund}, these limits can be avoided
if there is a strong convergence of the angular momentum
flux near the base of the CZ which would effectively confine
the equatorward return flow to a thin layer (this applies to
the return flow of several m s$^{-1}$ required for the operation
of the FT dynamo; weaker equatorward flows may exist outside
of this layer).  In particular, the lower limit on $V_c$ 
in Eq.\ (\ref{eq:vamp}) scales as $L_t^{1/2}$ where $L_t$ 
is the width of this layer.  Reducing $V_c$ enough to yield 
an $\eta_t$ of 10$^{11}$ cm$^2$ s$^{-1}$ would require that the 
equatorward return flow be confined to a layer no wider than 
\begin{equation}\label{eq:L_t}
L_t \sim \frac{r_m V_c^2}{\sqrt{2} V_m \left| \del {\cal L}\right|} \sim
1.7 ~ \mbox{Mm} \sim 2 \times 10^{-3} R ~~~.
\end{equation}
Here we have used Eq.\ (\ref{eq:vamp}) with
$V_c = 3 \eta_t/L_c$, $\eta_t = 10^{11}$ cm$^2$ s$^{-1}$, $L_c = 5.5$ Mm,
$V_m \sim $ 2 m s$^{-1}$, $\left| \del {\cal L}\right| \sim$
3 km s$^{-1}$, $\delta \sim \sqrt{2} L_t / r_m$ and $r_m = 0.7 R$.
Larger values of $L_c$ or $V_m$ would imply even stronger gradients
(smaller $L_t$).  Furthermore, since $V_c^2$ scales as $\epsilon^{-1}$
[see Eq.\ (\ref{eq:vamp})], a value of $\epsilon \sim 0.2$ as suggested
by convection models (\S\ref{sec:fund}) would reduce the estimate
of $L_t$ in Eq.\ (\ref{eq:L_t}) by a factor of five, to 340 km.
No current FT models employ such an extremely asymmetric meridional 
circulation profile and it is questionable whether efficient equatorward 
transport of strong toroidal flux concentrations could even occur in 
such a thin layer.

Even if such a thin layer were to exist near the base of the CZ, 
$V_c$ and $\eta$ would still have to be large enough in the upper CZ
to satisfy Eqs.\ (\ref{eq:delta2}) [with Eq.\ (\ref{eq:vamp})] 
and (\ref{eq:qlimit2}).  These suggest that $V_c$
must be at least 30 m s$^{-1}$ at $r \sim 0.95$, independent
of the deeper structure and amplitude of the meridional flow.
If $L_c \gtrsim $ 5.5 Mm then
this suggests $\eta_t$ should be at least $5.5 \times 10^{11}$ cm$^2$ s$^{-1}$ 
in the upper CZ for any model with solar-like mean flows.
This is in fact satisfied by many current FT models, but again,
this is a conservative estimate.  If $L_c \gtrsim $ 30 Mm and 
$\epsilon \sim 0.2$, this limit becomes much more stringent
at $\eta_t \gtrsim 6.7 \times 10^{12}$ cm$^2$ s$^{-1}$.
In the lower CZ, for $N_c = 1$, Eq.\ (\ref{eq:qlimit2})
suggests $\eta_t \gtrsim 1.6 \times 10^{11}$ cm$^2$ s$^{-1}$ 
for $L_c \gtrsim $ 5.5 Mm and $\epsilon \sim 1$ or 
$\eta_t \gtrsim 2 \times 10^{12}$ cm$^2$ s$^{-1}$ 
for $L_c \gtrsim $ 30 Mm and $\epsilon \sim 0.2$.
Multiple cells in radius ($N_c > 1$) could in principle help mitigate 
these limits on $V_c$, $\eta$, and $L_t$ but they are known to have 
an adverse effect on the operation of FT dynamos \citep{jouve07}.

Thus, we conclude that current advection-dominated FT dynamo models 
are not strictly self-consistent.  In particular, the amplitude of
turbulent transport they typically assume is not commensurate with the
mean flows they employ.  Again, this has been argued before based on
mixing-length theory and convection simulations but here we
demonstrate more generally that it is a direct consequence of 
Eq.\ (\ref{eq:bob}).   

A potential way out of this dilemma is by moving to the
diffusion-dominated regime.  This would entail using values of 
$\eta_t \gtrsim 10^{12}$ cm s$^{-1}$ throughout most of the convection zone.
Such models have indeed had some success in modeling the solar cycle,
and may actually do better in certain respects than
advection-dominated models.  Examples include reproducing the observed
cycle amplitude-period relationship and triggering grand minima
through variations in the meridional flow
\citep{jiang07,yeate08,karak10}.  However, it can be a challenge
for diffusion-dominated FT models (or any model in which the
transport by turbulent diffusion or magnetic pumping is very
efficient) to achieve cycle periods as long as 11 years.  As noted
above, efficient turbulent transport in a Babcock-Leighton dynamo
tends to short-circuit the region over which the dynamo operates and
to thus decrease the cycle period.  This is sometimes avoided by
allowing the meridional circulation to extend well below the
convection zone.  However, if the turbulent diffusivity there is 
low, then these models may suffer from the same problem described 
here (namely that the local amplitude of $\eta_t$ is incommensurate 
with the local amplitude of the meridional flow).  Alternatively,
one could place the turnover radius $r_0$ higher up in the CZ to
achieve a slower meridional flow in the lower CZ and thus a 
longer period (see Fig.\ 3). Still, the downward turbulent 
transport cannot be so efficient that it suppresses the poleward 
migration of residual flux from active regions that is seen 
in photospheric observations.

Alternatively, it has been argued that the quenching
of turbulent transport by Lorentz force feedbacks
may be able to salvage the advection-dominated
regime \citep{guerr09, munoz11}.
However, it remains to be seen whether or not this 
is viable.  Quenching would be most effective for strong
toroidal fields near the base of the CZ as opposed to
the relatively weak poloidal fields whose turbulent
transport across the CZ could potentially short-circuit 
the advection by the meridional flow.  In other words, 
quenching is more likely to regulate the first role noted
above for the meridional circulation in FT dynamos (equatorward
advection of toroidal flux) as opposed to the second (coupling 
of poloidal and toroidal sources), and it is the second which 
generally defines the advection-dominated regime.

Furthermore, the arguments presented here demonstrate 
that a significant reduction in $V_c$ by Lorentz force feedbacks 
as captured by quenching mechanisms would have substantial consequences 
for the mean flows.   For example, if we treat $V_\Omega$ as fixed, 
Eq.\ (\ref{eq:limit1}) indicates that a reduction in $V_c$ by 
an order of magnitude may be accompanied by a reduction 
of the meridional flow speed $V_m$ by two orders of magnitude
(unless the efficiency factor $\epsilon$ increases).
This follows from the concept of gyroscopic pumping 
discussed in \S\ref{sec:gp}, provided that the nature of the
Lorentz forces is solely to suppress convection.  However, 
if the Lorentz force exerts its own mean torques, which is
likely, then this should be included in the right-hand-side of 
Eq.\ (\ref{eq:bob}), with corresponding modifications
to Eqs.\ (\ref{eq:vamp}) and (\ref{eq:limit1}).  The time 
dependence should also be taken into account, potentially
mitigating limits on $\eta_t$ for a given meridional flow speed.
Even so, a self-consistent treatment of diffusivity quenching 
that takes into account its dynamical effect on the maintenance 
of mean flows could dramatically alter the operation of 
a FT dynamo by substantially modifying the meridional 
circulation and, to a lesser extent, the differential
rotation.

A related possibility is that turbulent transport at very high
magnetic Reynolds numbers may simply be less efficient than suggested
by crude, kinematic representations based on the concept of turbulent
diffusion.  Nonlinear processes such as the dynamical alignment of
fields and flows that contribute to dynamo saturation may also
suppress turbulent transport in a way that is more subtle than a
quenched local diffusion coefficient.  This might require more
sophisticated mean-field and/or MHD convection models to properly
capture.  Still another possibility, of course, is that the solar
dynamo does not follow the canonical
Babcock-Leighton/Flux-Transport paradigm.

\section{Summary}\label{sec:summary}

In this paper we estimate the amplitude and scale of the convective 
motions responsible for maintaining the solar differential rotation
and meridional circulation.  This estimate is based only on the
observed properties of the mean flows in the Sun together with
three fundamental physical premises grounded in the MHD equations.

The first and most important of these three premises is represented 
by Eq.\ (\ref{eq:bob}).  This tells us that, in order to achieve
a statistically steady state, the angular momentum transport by
the convective Reynolds stress must balance the advection of 
angular momentum by the meridional flow.  This provides a direct
link between the amplitude of mean flows and the amplitude of 
convective motions such that observations of the former can
set constraints on the latter.  Note that this dynamical balance
is to be understood in a time-averaged sense, filtering out 
solar cycle variations of meridional and zonal flows.

The second physical premise we rely on is that of mass conservation
[Eq.\ (\ref{eq:divmc}) in the Appendix].  This allows us to estimate 
the net equatorward flow speed in the deep CZ based on observations of 
the poleward flow near the surface as shown in Fig.\ \ref{fig:vtheta}.  
This result in turn allows us to extend the diagnostic power of our 
first premise to the deep convection zone where mean flows are 
established.

The third premise is that of thermal wind balance, Eq.\ (\ref{eq:twb}).
Although this is not used directly in our estimates for the convective
velocity amplitude $V_c$, it helps to justify the concept of gyroscopic 
pumping discussed in \S\ref{sec:gp}.  In particular, when coupled with 
helioseismic rotational inversions, it suggests that the principle 
mechanism responsible for maintaining the solar meridional circulation 
is the inertia of the differential rotation, as represented by the 
Coriolis force (\S\ref{sec:twb}).

Together with these three physical premises, we use three observational
foundations.  The first is the solar differential rotation inferred
from helioseismology.  The second is a persistent poleward meridional
flow from low to high latitudes for $r \gtrsim 0.95 R$ with an amplitude
of roughly 10--15 m s$^{-1}$.  The third observational foundation is 
the mean density profile $\rho(r)$ in the CZ.  Although this is obtained 
from a solar structure model (Model S), we regard it as an observation 
because it is verified to within a few percent by helioseismic structure 
inversions.

The results suggest that the amplitude of convective velocities in the
upper CZ ($r \sim 0.95 R$) must be at least 30 m s$^{-1}$ in order to
sustain the observed mean flows.  Analogous limits in the lower CZ
are less reliable due to uncertainties about the meridional flow 
but reasonable inferences suggest that convective amplitudes can
be no less than about 8 m s$^{-1}$ at $r \sim 0.75$.

The existence of the near-surface shear layer (NSSL) provides a smoking 
gun that we can exploit to link these convective amplitudes to a size
scale.  In particular, it suggests that the characteristic Rossby 
number crosses unity somewhere near the base of the NSSL at 
$r \sim 0.95 R$ (MH11).  Together with the velocity limits obtained
in \S\ref{sec:amp}, this implies deeper convective motions can
be no smaller than 5.5 Mm.  This is of course a conservative limit.  
Surface convection simulations suggest that the size scale of 
convective motions progressively increases with depth, from 
the $\sim $ 1 Mm granulation scales at the surface to
$\sim $ 30 Mm supergranulation scales by $r \sim 0.97 R$,
due mainly to an increase in the density and pressure scale
heights \citep[e.g.][]{rempe11,tramp11}.  Still, the limit
given here provides an independent constraint on the size of
deep convection that depends only on the three physical
premises and observational foundations noted above
(along with a fourth premise that $R_o \lesssim 1$ marks
the base of the NSSL).

The lower limits on convective velocity and size scales reported 
here are consistent with the upper limits inferred from local
helioseismology by \cite{hanas10} provided that the characteristic
size of the convective motions at $r \sim 0.95 R$ lies somewhere in 
the range between 5.5--83 Mm.  This corresponds to a range in spherical 
harmonic degree of $50 \lesssim \ell \lesssim 750$.  This range may be 
too wide to provide a strong constraint on convection models but it can 
be narrowed somewhat if one requires that motions at and below $r \sim
0.95 R$ be no smaller than supergranules, so $L_c \gtrsim 30 $ Mm and
$\ell \lesssim 130$.

Since our work is concerned specifically with the relationship between
convective transport and mean flows, it has important implications
for Flux Transport dynamo models (\S\ref{sec:ftd}).  In particular,
it suggests that advection-dominated FT models are not self-consistent
in the sense that the assumed magnitude of convective transport
is generally too low to account for the mean flows they require to 
operate.  This has been argued before based on estimates of $V_c$ and $L_c$
obtained from mixing length theory and convection models (cf.\ Fig 3).  
However, here we demonstrate that it is a more general consequence of 
the need to sustain the mean flows against their own inertia.

The essence of the problem can be appreciated simply from Eq.\ (\ref{eq:limit1}).  
If $(2 \delta)^{1/2} \sim 1$ and $V_\Omega \gg V_m$, then $V_c$ must be significantly 
larger than $V_m$.  Thus, one would expect convection to dominate transport
over the meridional circulation unless its spatial and temporal correlation 
scales are small.  However, these correlation scales cannot be too small because 
if they were, the influence of rotation on the convection would not be strong
enough to establish a solar-like differential rotation 
(i.e.\ $R_o$ would be greater than unity; see \S\ref{sec:himp}).  

This result casts some doubt on the advection-dominated flux-transport 
paradigm as a viable model of the solar cycle but it does not 
necessarily rule it out.  An alternative possibility is that convective 
flux transport in the extreme parameter regimes of the solar
interior is much less efficient than suggested by turbulent diffusion
and that we have much to learn about reliably representing
this transport in mean-field dynamo models.

The estimates reported here are independent of mixing-length theory
and convection simulations but they give compatible values for $V_c$
and $L_c$, suggesting internal consistency (\S\ref{sec:models}).
Furthermore, they suggest that the velocity and length scales
responsible for maintaining solar mean flows cannot be drastically
different from those currently exhibited by global convection
simulations.  This bodes well for the future; as global convection models
continue to move toward higher resolution and improved representations
of the upper and lower boundary layers, they should be able to capture
the essential physics underlying the solar differential rotation and
meridional circulation with increasing fidelity.

\acknowledgements
We thank Yuhong Fan and Bidya Karak for comments on the manuscript and the anonymous 
referee for constructive criticisms that have improved the presentation.
We also thank Rachel Howe for providing the rotational inversions used to construct 
Fig.\ \ref{fig:Lfig}, and Kyle Augustson, Ben Brown, Nick Nelson, and Juri Toomre 
for many enlightening and inspiring discussions.  This work is supported by 
NASA grants NNH09AK14I (Heliophysics SR\&T) and NNX08AI57G (Heliophysics 
Theory Program).  The National Center for Atmospheric Research is sponsored 
by the National Science Foundation.


\appendix

\subsection{Baroclinicity as a Source of Differential Rotation}\label{baroshear}

In \S\ref{sec:twb} we argued that baroclinic forcing cannot account
for the observed sense of the solar meridional flow (poleward near
the surface).  In this Appendix we demonstrate further that 
baroclinicity alone cannot account for the solar differential 
rotation as inferred from helioseismology. Similar issues
have been studied for decades within the context of planetary
and stellar atmospheres \citep{elias51,read86,tasso78}.

The importance of baroclinicity for shaping the solar differential
rotation profile is undisputed.  Theoretical models, mean-field 
models, and global convection simulations all suggest that 
baroclinic forcing is necessary to account for the conical nature
of mid-latitude $\Omega$ surfaces in the solar CZ inferred from 
helioseismology \citep{kitch95,robin01,brun02,rempe05,miesc06,balbu09}.
However, this result should not be misinterpreted to attribute
the solar differential rotation entirely to baroclinic forcing.

Such a clarification is particularly timely in light of the recent 
work by \cite{balbu12}.  They demonstrate that the centrifugal 
distortion of the base of the convection zone can produce thermal 
gradients along isobaric surfaces that can in turn induce differential
rotation through baroclinic forcing.  This background shear may
then interact with convection to produce the observed mean flows.
We do not dispute this argument but again we caution against 
a potential misinterpretation of their results. In particular, we 
argue that baroclinicity alone cannot induce a net equatorward 
angular velocity gradient 
($\pd \Omega / \pd \theta > 0$ in the northern hemisphere, 
or NH) throughout the CZ as exhibited by helioseismic rotational 
inversions.

Our argument begins with the time-dependent zonal vorticity equation
(\ref{eq:meridional}).  Again we neglect the Reynolds stress, 
Lorentz force, and viscous diffusion.
We proceed to consider a fixed entropy gradient 
$\pd \left<S\right>/\pd \theta$ 
and we ask what differential rotation such baroclinic forcing will 
produce when subject to a given initial condition 
$\Omega(r,\theta,t=0) = \Omega_i(r,\theta)$ and $\vv_m(t=0) = 0$.

As noted in \S\ref{sec:twb}, a poleward entropy gradient will
induce a clockwise meridional circulation cell in the NH 
($\pd \left<S\right>/\pd \theta < 0$, 
$\left<\omega_\phi\right> > 0$).  Although it was not necessary
for the arguments presented in \S\ref{sec:meanflows}, we now adopt 
the anelastic approximation so the mean mass flux is divergenceless:
\begin{equation}\label{eq:divmc}
\dv \left<\rh \vv_m\right> = 0
\end{equation}
where $\rh(r)$ is the background density profile, averaged over 
latitude, longitude, and time.  Thus, the conservation of mass 
requires that the generation of zonal vorticity 
$\left<\omega_\phi\right>$ be associated with flows both toward
and away from the rotation axis.  This is crucial in understanding
the significance of the results that follow.

In order to address the generation of rotational shear, we must consider
the time-dependent version of the zonal momentum Eq.\ (\ref{eq:gp}).
Another crucial realization required to appreciate the results that follow
is that there is no baroclinic component of the net axial torque
${\cal F}$.  Recall that ${\cal F}$ involves only the Reynolds 
stress, the Lorentz force, and the viscous diffusion (explicit
expressions are given in MH11).  If baroclinic forcing is to
induce a differential rotation, it must do so by means of the
meridional flow.  Thus, we can set ${\cal F} = 0$, which yields
(MH11)
\begin{equation}\label{eq:dLdt}
\rh \frac{\pd {\cal L}}{\pd t} = - \left<\rh \vv_m\right> \bdot \del {\cal L} ~~~.
\end{equation}
In Eq.\ (\ref{eq:dLdt}) we have again used the anelastic approximation
so we have replaced $\rho$ with $\rh$.

Two things are immediately apparent from Eq.\ (\ref{eq:dLdt}).
First, for a given meridional flow $\left<\rh \vv_m\right>$, this
is identical to the equation for passive advection of a scalar field.  
Second, the only steady solution is one in which 
$\left<\rh \vv_m\right> \bdot \del {\cal L} = 0$.  Such a steady
state can be achieved in one of two ways.  Either the meridional
flow must vanish $\left<\vv_m\right> = 0$ or the specific 
angular momentum ${\cal L}$ must be constant on streamlines.
The latter case, ${\cal L} = \lambda^2 \Omega$ constant on 
streamlines, would necessarily be associated with an anti-solar 
differential rotation profile, such that the poles would spin 
faster than the equator ($\pd \Omega / \pd \theta < 0$ in the NH).
We will return to the former case, that of $\left<\vv_m\right> = 0$
below.

Now return the time-dependent problem outlined above in which we
follow the response of the mean flows to a specified baroclinic 
forcing $\pd \left<S\right>/\pd \theta$.  Let us assume for 
simplicity that the initial angular velocity profile is 
cylindrical, so $\Omega_i = \Omega_i(\lambda)$.  As noted 
below, this is not a necessary assumption but it serves well
to illustrate the main point.  Furthermore, we will assume that
the initial angular momentum gradient is directed away from
the rotation axis, so $d{\cal L}_i/d\lambda > 0$, where
${\cal L}_i = \lambda^2 \Omega_i$.  This is true for the 
current Sun, it is true for a uniform rotation ($\Omega_i = \Omega_0$),
and we suspect that it is true in general for stars since the
alternative, $d{\cal L}_i/d\lambda < 0$ is unstable according
to the Rayleigh criterion \citep{tasso78}.  Note that the
case $\del {\cal L} = 0$ is a fixed point; if this were the 
initial state, meridional circulation would not influence the 
zonal flow and baroclinicity would not induce a differential 
rotation.

Consider a volume ${\cal V}$ defined by $r \leq R$ and 
$\lambda \leq \lambda_0$, where $R$ is the solar surface and
$\lambda_0$ is a fiducial cylindrical radius that can lie 
anywhere between zero and $R$.  Integrating Eq.\ (\ref{eq:dLdt})
over ${\cal V}$ and using Eq.\ (\ref{eq:divmc}) yields
\begin{equation}\label{eq:dLVdt}
\frac{d L_{\cal V}}{dt} =
\int_{\cal S} {\cal L} \left<\rh \vv_m\right> \bdot d\ssurf =
2 \pi \lambda \int_{z_-}^{z_+} {\cal L} \left<\rh v_\lambda\right> ~ dz 
\end{equation}
where 
\begin{equation}
L_{\cal V} = \int_{\cal V} \rh {\cal L} ~ d{\cal V} 
\end{equation}
is the total angular momentum within ${\cal V}$
and $z_{\pm} = \pm  (R^2 - \lambda_0^2)^{1/2}$.  We have assumed
that there is no flow across $r = R$.  

At the initial time $t = 0$ the $\Omega$ profile is cylindrical
so ${\cal L}$ can be pulled out of the integral on the 
right-hand-side (RHS).  Mass conservation then implies that 
the RHS vanishes and the rate of change of $L_{\cal V}$
vanishes.  However, the analogy with the passive scalar noted
above makes it clear that this cannot hold indefinitely.
Our intuition tells us that if $d {\cal L}_i/d\lambda > 0$
then the advection of angular momentum into our volume must
increase $L_{\cal V}$ over time.

This apparent inconsistency can be resolved if we consider a 
Taylor expansion at early times
\begin{equation}\label{eq:Taylor}
L_{\cal V}(t) = L_{\cal V}(0) + 
t ~ \left.\frac{d L_{\cal V}}{d t}\right|_{t=0} + t^2 ~ 
\left.\frac{d^2 L_{\cal V}}{d t^2}\right|_{t=0} + \ldots
\end{equation}

Applying a time derivative to (\ref{eq:dLVdt}) and substituting in
(\ref{eq:dLdt}) yields
\begin{equation}\label{eq:ddL}
\left.\frac{d^2 L_{\cal V}}{dt^2}\right|_{t=0} =  
2 \pi \lambda \int_{z_-}^{z_+} \left<\rh v_\lambda\right> \frac{\pd {\cal L}}{\pd t} dz =
2 \pi \lambda \int_{z_-}^{z_+} \rh \left<v_\lambda\right>^2 \frac{d {\cal L}_i}{d \lambda} ~ dz ~~~.
\end{equation}
In obtaining this result we have neglected the time dependence of $\left<\vv_m\right>$.
Thus, the second derivative of $L_{\cal V}$ is positive definite at $t = 0$.  Together 
with (\ref{eq:Taylor}) this implies that the angular momentum within ${\cal V}$ will 
increase with time.

Put another way, this implies that any persistent meridional circulation 
will tend to spin up the poles relative to the equator.  This is true
regardless of the (nonzero) amplitude or profile of the circulation and 
regardless of the source of the circulation, whether it be baroclinic
in nature or due to some other meridional forcing.  Furthermore, it holds
for any arbitrary initial ${\cal L}$ profile that is stable in the sense 
that $\pd {\cal L}_i/\pd \lambda > 0$.  In this general case, ${\cal L}$
surfaces would still intersect $R$ at two locations and ${\cal V}$ would
be defined relative to the surface ${\cal L} = {\cal L}_0$ such that 
$r \leq R$ and ${\cal L} \leq {\cal L}_0$.  Then Eq.\ (\ref{eq:ddL}) 
would still hold with  $d {\cal L}_i/d\lambda$ replaced by 
$\vert \del {\cal L}\vert$, $\left<v_\lambda\right>^2$ by 
$(\left<\vv_m\right> \bdot \uvn)^2$ (where $\uvn$ is the unit vector 
normal to ${\cal S}$) and the integration would proceed
over the surface ${\cal S}$.  The early time dependence would
still be given by (\ref{eq:Taylor}) with 
$(d L_{\cal V}/d t)_{t=0} = 0$ and $(d^2 L_{\cal V}/d t^2)_{t=0} > 0$. 

This argument can also be readily generalized to a spherical annulus.
If there is a radius $r_b$ below which the meridional flow 
$\left<\vv_m\right>$ becomes negligible, then one can define the
volume ${\cal V}$ to be bounded from above and below by $R$
and $r_b$, extending poleward of an ${\cal L}$ isosurface that 
extends from $R$ to $r_b$ (confined either to the northern or 
southern hemisphere).  Then the conclusion is unaltered; a 
baroclinically-driven meridional flow will increase the 
total angular momentum in the polar cap ${\cal V}$ over time, 
provided only that $\pd {\cal L}_i / \pd \lambda > 0$.

Spin-up of the poles can be avoided if the Rossby number is small,
which is a good approximation for the Sun.  Here the relevant 
Rossby number is that based on the meridional flow
$R_o^m = V_m / (2 \Omega_0 R$).  For $R_o^m \ll 1$, meridional flows
are redirected into zonal flows essentially immediately (on the
time scale of a rotation period), so the steady-state version 
of Eq.\ (\ref{eq:dLdt}) can be satisfied with 
$\left<\vv_m\right>=0$.  At the same time, a differential 
rotation will be quickly established that obeys thermal
wind balance, Eq.\ (\ref{eq:twb}).

This is indeed a viable way to generate an axial shear
$\pd \Omega/\pd z$ and may play an important role in the Sun.
However, it cannot spin up the equator relative to the
poles, as seen in the solar convection zone.  In the
limit $R_o^m \ll 1$, there is no angular momentum transport
across $\lambda$ surfaces.  Rather, at a given cylindrical
radius $\lambda$, any deceleration of $\Omega$ for some
range of $z$ must be balanced by an acceleration at some
other $z$.  This follows directly from the conservation of 
mass and angular momentum as outlined above.

This implies that a solar-like differential rotation 
characterized by an equatorward $\del \Omega$ in the
CZ can only be produced by baroclinic forcing if there
is a corresponding poleward $\del \Omega$ somewhere
below the CZ.  We cannot rule this out from helioseismic
rotational inversions but there is currently no evidence
for it.

We emphasize that the thermal wind equation (\ref{eq:twb}) 
possesses a geostrophic degeneracy in that a given 
entropy gradient $\pd\left<S\right>/\pd\theta$ is compatable
with an infinite number of $\Omega$ profiles, some solar-like
($\pd \Omega /\pd \theta > 0$ in the NH) and some anti-solar.
This arises because one can take a solution of this 
equation $\Omega_s(r,\theta)$ and add an arbiratry cylindrical
component $\Omega^\prime(\lambda)$ and the sum is also a
solution.  We argue that the convective Reynolds stress is
necessary to break this degeneracy and establish a solar-like
$\Omega$ profile.

In any case, it is clear that there must be some non-baroclinic 
forcing in the solar CZ in order to account for the observed mean
flows.  As noted in \S\ref{sec:intro}, solar observations indicate
a persistent poleward circulation in the solar surface layers
and mass conservation requires a net equatorward flow deeper 
down.  Since ${\cal L}$ isosurfaces are nearly cylindrical
throughout the CZ (MH11), this necessarily implies flow across
${\cal L}$ contours that must be balanced in a steady state
with a nonzero ${\cal F}$, as expressed in Eq.\ (\ref{eq:gp}).

\end{document}